\documentclass[lettersize,journal]{IEEEtran}
\usepackage{xcolor}
\usepackage{amsmath,amsfonts}
\usepackage{algorithmic}
\usepackage{algorithm}
\usepackage{array}
\usepackage[caption=false,font=normalsize,labelfont=sf,textfont=sf]{subfig}
\usepackage{textcomp}
\usepackage{stfloats}
\usepackage{url}
\usepackage{verbatim}
\usepackage{graphicx}
\usepackage{cite}
\begin{document}

\title{Analog Solver Circuit for Linear Symmetric Positive-Definite Systems at a Complexity Independent of Matrix Size}


\author{
    Osama Abdelaleim, Arun Prakash, Ayhan Irfanoglu, Veljko Milutinovic
    \thanks{Osama Abdelaleim (email:oabdelal@purdue.edu), is a doctoral candidate at the Lyles School of Civil and Construction Engineering, Purdue University, West Lafayette, Indiana, USA}    
    \thanks{Arun Prakash and Ayhan Irfanoglu are with the Lyles School of Civil and Construction Engineering, Purdue University, West Lafayette, Indiana, USA}
    \thanks{Veljko Milutinovic is with the School of Electrical Engineering, University of Belgrade, Belgrade, Serbia}

}

\markboth{IEEE Transactions on Circuits and Systems I: Regular Papers}
{Osama Abdelaleim, Arun Prakash, Ayhan Irfanoglu, Veljko Milutinovic,
\MakeLowercase{\textit{(et al.)}:Near instantaneous O(1) Analog Solver Circuit for Linear Symmetric Positive-Definite Systems}}


\maketitle

\begin{abstract}
Accelerating the solution of linear systems of equations is critical due to their central role in numerous applications, such as numerical simulations, data analytics, and machine learning. 
This paper presents an analog solver circuit designed to accelerate the solution of symmetric positive definite (SPD) linear systems of equations. 
The proposed design leverages non-inverting operational amplifier configurations to create a negative resistance circuit, effectively modeling any symmetric system. 
The paper details the principles behind the design, optimizations of the system architecture, and numerical results that demonstrate the robustness of the design. 
The findings reveal that the proposed system solves symmetric diagonally dominant (SDD) matrices with $O(1)$ complexity, achieving the theoretical maximum speed as the circuit relies solely on resistors. 
For non-diagonally dominant SPD systems, the solution speed depends on matrix properties, specifically eigenvalues and diagonal dominance deviation, but remains independent of the size of the matrix.
\end{abstract}



\section{Introduction}\label{sec:Intro}

\IEEEPARstart{S}{olving} systems of linear equations are fundamental for addressing real-world challenges across various domains, especially in this era of big data where the volume and complexity of datasets continue to grow exponentially. 
Although powerful, traditional digital computing approaches are encountering limitations in parallelization and scalability, particularly as Moore's Law approaches its asymptotic plateau. 
Accelerating the solution of linear systems of equations will be immensely impactful for applications such as scientific simulations, data analytics, and machine learning.

\subsection{Existing Paradigms for Solving Large Systems of Equations}
\label{sec:existing}

For decades, the primary engine for solving systems of linear equations has been digital computing. 
Classical direct factorization methods, such as Gaussian elimination, LU decomposition, and Cholesky factorization, offer exact solutions but typically exhibit a computational complexity of $O(n^3)$, where $n$ is the number of unknowns \cite{golub2013}.
Iterative methods, including Jacobi, Gauss-Seidel, Successive Over-Relaxation (SOR), and Conjugate Gradient (CG), offer alternatives, particularly for large, sparse systems \cite{saad2003}.
While often exhibiting lower complexity $O(n^2)$ per iteration, the convergence rate of iterative methods depends heavily on matrix properties such as condition number \cite{saad2003}.
As the size ($n$) of the systems are ever increasing in modern applications, digital solvers, both direct and iterative, face significant bottlenecks.
Executed on traditional von Neumann architectures, they have costly time and energy overheads associated with data movement between memory and processing units, pushing power consumption and execution time to prohibitive levels \cite{compEnergy}.
This challenge motivates the exploration of fundamentally different computing paradigms such as quantum computing and analog computing.

With quantum computing, algorithms like the Harrow-Hassidim-Lloyd (HHL) algorithm theoretically promise an exponential speedup for solving certain linear systems, potentially achieving solver complexity of $\text{log}(n)$ under specific conditions on the system matrix $A$ and the desired output precision\cite{quantum}.
In such approaches, obtaining explicit solution components involves a separate readout process with linear complexity $O(n)$. 
Although its long-term promise holds, quantum computing faces challenges such as qubit coherence times and error rates.
Thus, it is currently not a viable option for large-scale problems \cite{preskill2018quantum,quantum2}.

An alternative gaining significant attention is analog computing.
Advances in materials science and device physics have led to the development of analog Compute-In-Memory (CIM) architectures \cite{sebastian2020memory}.
This paradigm shifts computation directly onto memory elements, drastically reducing data movement\cite{ielmini2018memory}.
Advancements of non-volatile resistive memory technologies, particularly memristors (or Resistive RAM - ReRAM), organized into crossbar arrays led to the evolution of this architecture \cite{strukov2008missing}. 
Memristors are variable resistor devices that can act as memory by changing their conductance values.
Through ongoing research, the number of unique conductance values that a single memristor device can represent has progressively increased, with milestones demonstrating 2 bits \cite{mem2bits}, 5 bits \cite{mem5bits}, 9 bits \cite{mem9bits}, and ultimately 11 bits \cite{mem11bits}. 
The precision of these unique values is limited by reading noise, writing variabilities, conductance drift, retention failure, and programming inaccuracies \cite{memerror1,memerror2,memerror3,memerror4,memerror5,memerror6}.
Techniques to overcome these challenges allowed a progressive increase in precision. 
Use of an error compensation technique across multiple (no more than five) arrays achieved double precision \cite{song2024programming}. 
These advancements highlight the potential of analog computing to perform computations with accuracy comparable to digital computing.

Analog computing systems, particularly those leveraging memristor-based compute-in-memory architectures, have shown promise in performing matrix-vector multiplication (MVM) efficiently, potentially reducing computational complexity and energy consumption \cite{li2018analogue}.
The crossbar arrays in the architecture naturally implement MVM, a core operation in many linear system solvers, by leveraging fundamental physical principles like Ohm's Law and Kirchhoff's current law \cite{hu2018memristor}.
The complexity of MVM operation using memristors in the crossbar architecture is $O(1)$ compared to $O(n^2)$ in digital computing \cite{sebastian2020memory}. 
Researchers report that a MVM CIM circuit can achieve 100x better energy efficiency than GPUs \cite{invitedtut}.

Solving linear systems ($Ax=b$) using analog CIM has consequently gained a lot of attention,  precisely because it relies heavily on MVM and demands high precision.
These analog designs can be grouped into three classes: hybrid solvers, dynamical solvers, and feedback solvers. 
The hybrid (digital + analog) systems solve linear equations iteratively using MVM circuits. 
Progress in hybrid systems is observed in the increasing integration of analog components within digital computers: ranging from the use of analog CIM as a quick generator of an initial guess for a digital solver \cite{richter2015memristive}, to the implementation of analog preconditioners \cite{precond}, and to accelerating all MVM operations within conventional iterative algorithms such as the generalized minimum residual (GMRES) \cite{le2018mixed}. 
These systems show potential in accelerating the MVM part of algorithms by reducing run time by up to 1500x and saving energy by 8.5x compared to digital computing \cite{richter2015memristive}.

The second class of analog designs is dynamical systems, which convert the system of linear equations to a system of differential equations (DEs) whose steady-state solution is the solution of the linear system of equations. 
Analog systems are transient systems best described using partial differential equations (PDEs) and ordinary differential equations (ODEs). 
Researchers use analog integrator circuits to solve the equivalent system of DEs \cite{ulmann2019solving,huang2016evaluation}, while other researchers utilize transconductance devices in their designs to solve the DEs \cite{hasler2020continuous,natarajan2020analog}. 
To improve scalability, recent work has also proposed integrator circuits based on ring oscillators, which can reduce both area and power consumption compared to conventional operational amplifier (OA)-based integrators \cite{li2025voltage}.
The main drawback of these systems is that, by converting the linear system into DEs, the solution can have a long settling time as it depends on dynamic components such as operational amplifiers and capacitors.

The third class of analog designs is the feedback methods. 
These systems model iterative algorithms into feedback networks.
The goal of these systems is to achieve one-step solutions where the solution time is independent of the size of the system. 
Using the MVM crossbar architecture in feedback with Operational Amplifiers (OAs) can solve positive definite (PD) matrices in one step \cite{sun2019solving}. 
For sparse systems, the direct CIM design is shown to have a complexity that is independent of the matrix size ($n$) and given by $O(1/\lambda_{min})$ where $\lambda_{min}$ denotes the minimum eigenvalue of the matrix\cite{sun2019fast}. 
In comparison, the complexity of solving sparse matrices by digital computing using Conjugate Gradient method is $O(n)$ and by quantum computing it is $O(\kappa^2\text{log}(n))$ where $\kappa$ denotes the condition number of the matrix \cite{sun2019fast}. 
For more general matrices, complexity of the feedback analog computing is independent of matrix size but quadratic in the condition number ($O(\kappa^2)$) \cite{mannocci2021fully}. 
In addition, these systems accelerate computing the inverse of the system matrix at a complexity of $O(n^2)$ compared to a complexity of $O(n^3)$ for digital computing \cite{lin2025memristive}. 
One notable advantage of these analog designs is that they allow the use of the same architecture for multiple matrix operations \cite{mannocci2023generalized}.
To achieve higher speed and reduce the physical space of the design, researchers aim to reduce the number of active components needed in the design \cite{luo2025smaller}. 
However, these designs always contain active components that can still result in long settling time, which is problem-dependent.

A common characteristic across many of these analog solver architectures, particularly those relying on dynamical systems or feedback loops incorporating active components, is the presence of inherent settling times. 
Whether due to explicit time constants (RC) in dynamical systems, the propagation delays, the finite gain-bandwidth product of OAs used in feedback and peripheral circuits, or the convergence time of the implemented algorithm itself, analog computation requires long settling time to converge to the final solution. 
This settling time can depend on the system size, matrix properties (such as condition number), and circuit non-idealities, potentially limiting the achievable speedup, especially compared with the near-instantaneous potential of passive MVM.

\subsection{Novel Approach Using Configurable Network of Resistors}
\label{sec:novelapproach}

Symmetric positive definite (SPD) linear systems arise in a wide range of scientific and engineering computations. 
Prominent examples include finite element solvers for solid mechanics, optimization of convex problems, and covariance-driven computations in filtering and uncertainty quantification. 
Because these applications often require solving large SPD systems repeatedly, accelerating SPD solvers is critical.

Here, we introduce a fundamentally different analog CIM design to solve symmetric positive definite (SPD) linear systems of equations at unprecedented speed. 
Instead of modeling the mathematical process of solving a linear system of equations, the proposed analog design is constructed to directly map the given linear system of equations $Ax=b$ into a resistive network.
The configuration and conductance values of this network are derived directly from the matrix $A$ and the vector $b$. 
The governing equations of the resistive network in this design are the given linear equations, and the set of nodal voltages in the network is the solution vector of the given linear system.

To illustrate, first, a preliminary circuit design based on direct mapping is developed to solve a $n\times n$ symmetric positive-definite (SPD) matrix system of equations.
However, the speed and accuracy of the solution are limited by the number of active positive off-diagonal entries (up to $(n^2-n)/2$ entries).
The reason is that the circuit resulting from mapping the positive off-diagonal terms contains active components, namely OAs in a negative feedback loop, whereas the negative off-diagonal terms are mapped directly into passive resistors.
The proposed design transforms the $n\times n$ SPD system to a $2n\times2n$ SPD system with at most $n$ positive off-diagonal values without adding extra computations.
Solving the transformed system is shown to require fewer components in the circuitry, speed up the convergence by orders of magnitude, consume less power, and allow the design to utilize the MVM architecture.

For the class of Symmetric Diagonally Dominant (SDD) systems, the mapping of the transformed system results in a purely passive resistive network, where the solution is achieved at a near-instantaneous $O(1)$ rate. 
For the broader SPD class, the mapping of the transformed system results in a resistive network that includes active components, particularly OAs in negative feedback.
The speed and accuracy of the solution in this case depend on the properties of the matrix $A$, such as the smallest eigenvalue, but, critically, remain independent of the size of the matrix.


The article is divided into five sections. 
Section \ref{sec:Background} presents the underlying theory of representing a linear system of equations with an equivalent network of resistors.
Section \ref{secDes1} details the preliminary design of the analog circuit to solve $n$ unknowns along with its performance in terms of stability and settling time. 
Section \ref{secDes2} presents the proposed design that models $2n$ unknowns detailing improvements in speed, number of circuit components, and power consumption without affecting stability. 
Finally, Section \ref{sec:conc} summarizes the advantages and limitations of the proposed design.


\section{Background}\label{sec:Background}

The proposed design aims to construct a circuit comprising a network of resistors corresponding to an $n$x$n$ linear symmetric positive-definite system of equations:
\begin{equation}
A\, x=b
\end{equation} 
We first present how such a system of equations may be represented as a circuit and then discuss the necessary modifications to it to account for the presence of negative resistance values.

\subsection{Equivalent Resistive Network}\label{sec:EqResNet}

A symmetric system of equations can be modeled using weighted undirected graphs and resistive networks \cite{GrembanPHD}. 
In this representation, the network comprises $n$ nodes, where each element of the unknown vector $x$ corresponds to the voltage at a node, and each element of the known vector $b$ represents the external current supplied to the corresponding node. 
The matrix $A$ represents the conductance matrix, assembled from the conductance values of the resistors that connect the nodes and the resistors that connect the nodes to the ground.

\begin{figure}[t]  
\centering 
\includegraphics[page=1,width=0.5\textwidth]{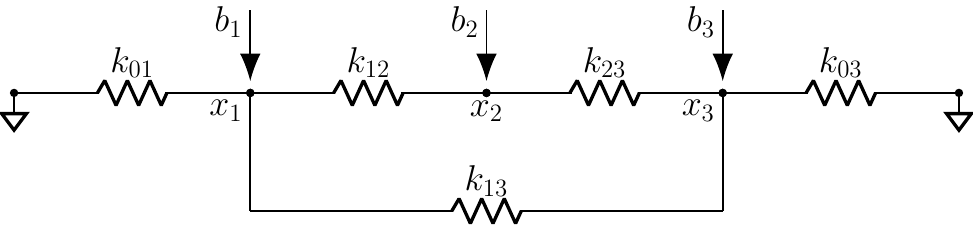} 
\caption{A resistive network example for a 3x3 system $Ax=b$. The unknown vector $x$ represents the voltage at the nodes, while the right hand side vector $b$ represents the external current going in to the nodes. The matrix $A$ represents the conductance of the resistors between the nodes.}
\label{fig:EquSysEx} 
\end{figure}

Consider the example shown in Figure \ref{fig:EquSysEx}. 
The system of equations corresponding to the circuit shown in Figure \ref{fig:EquSysEx} is the following 3x3 system:
\begin{multline}
\begin{bmatrix}
k_{01}+k_{12}+k_{13} & -k_{12} & -k_{13}\\
-k_{12} & k_{12}+k_{23} & -k_{23}\\
-k_{13} & -k_{23} & k_{03}+k_{13}+k_{23}\\
\end{bmatrix}
\begin{Bmatrix}
x_1 \\ x_2 \\ x_3
\end{Bmatrix}
\\=
\begin{Bmatrix}
b_1 \\ b_2 \\ b_3
\end{Bmatrix}
\label{eqn:3x3sys}
\end{multline}
Note that, for a resistor $k_{ij}$ connected to nodes $i$ and $j$ in the circuit, Ohm's law governs the current flowing out of these nodes and through the resistor as:
\begin{equation}
\begin{bmatrix}
k_{ij} & -k_{ij}\\
-k_{ij} & k_{ij}
\end{bmatrix}
\begin{Bmatrix}
x_i \\ x_j
\end{Bmatrix}
=
\begin{Bmatrix}
i_i^{el} \\ i_j^{el}
\end{Bmatrix}
\label{eqn:current_i_j}
\end{equation}
where $k_{ij}$ is the conductance of the resistor between nodes $i$ and $j$, $x_i$ and $x_j$ are the voltage at nodes $i$ and $j$, respectively, and $i_i^{el}$ and $i_j^{el}$ are the current flowing out of nodes $i$ and $j$, respectively. 
Similarly, the current $i^{el}_i$ flowing out of node $i$ due to a resistor connected between node $i$ and the ground is obtained as follows:
\begin{equation}
k_{0i}\ x_i=i_i^{el}
\label{eqn:current_i_gnd}
\end{equation} 
where $k_{0i}$ is the conductance of the resistor between node $i$ and the ground. 
The resistive network satisfies Kirchhoff’s current law (KCL), requiring that the summation of the current flowing out of each node $i$ be equal to the external current $b_i$ flowing into the node. 
Assembling the resistor Equations~\eqref{eqn:current_i_j}-\eqref{eqn:current_i_gnd}, for each resistor in the network, one obtains the system in Equation~\eqref{eqn:3x3sys}.
This system is easily solved by simply measuring the voltage at all the nodes that represent the unknowns $x_i$. 

\begin{figure}[!t]  
\centering 
\includegraphics[page=2,width=0.5\textwidth]{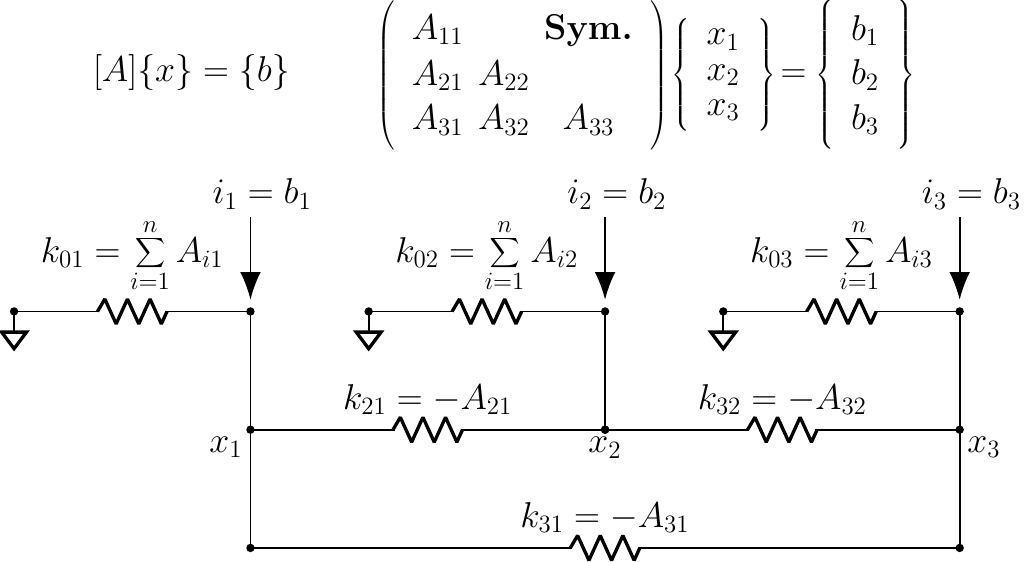} 
\caption{
Theoretical construction of a resistive network for a given 3x3 system $Ax=b$.
The unknown vector $x$ represents the voltage at the nodes, while the right hand side vector $b$ represents the external current going into the nodes. 
Conductance of the resistors are obtained from the components of $A$.
}
\label{fig:Ax=b} 
\end{figure}

In this formulation, the off-diagonal terms of matrix $A$ are the negative of the conductance values between the corresponding nodes, and the sum of each column of the matrix $A$ is the conductance value of the resistor connecting the corresponding node to the ground, i.e.
\begin{align}
k_{ij} &= -A_{ij}  \>\>\>\>\>\>\> for \>\> i\neq j   \label{equ:kij} \\ 
k_{0i} &= \sum_{j=1}^{n} A_{ji}  \label{equ:ki0}
\end{align}
where $A_{ij}$ is the element at row $i$ and column $j$ of the given matrix $A$. 
Thus, given a symmetric system of equations $Ax=b$, the equivalent resistive network can be constructed by computing the conductance of the resistors in circuit using Equations \eqref{equ:kij}-\eqref{equ:ki0}.
For example, a given 3x3 symmetric system of equations $Ax=b$ can be represented by a resistive network as shown in Figure \ref{fig:Ax=b}. 
Note that the conductance of the resistors resulting from Equations \eqref{equ:kij}-\eqref{equ:ki0} can either be negative or positive. 
A positive conductance is represented with a standard resistor, while the negative conductance is represented by the negative resistance circuit shown in Section \ref{secNegRes}.

\subsection{Negative Resistance Circuit}\label{secNegRes}
Negative resistance can be realized by designing a circuit that reverses the direction of current flow between two nodes \cite{chua1983negative}. 
Here, we extend the traditional non-inverting operational amplifier configuration to reverse the current between any two nodes.

\begin{figure}[!t]  
\centering 
\includegraphics[page=3,width=0.5\textwidth]{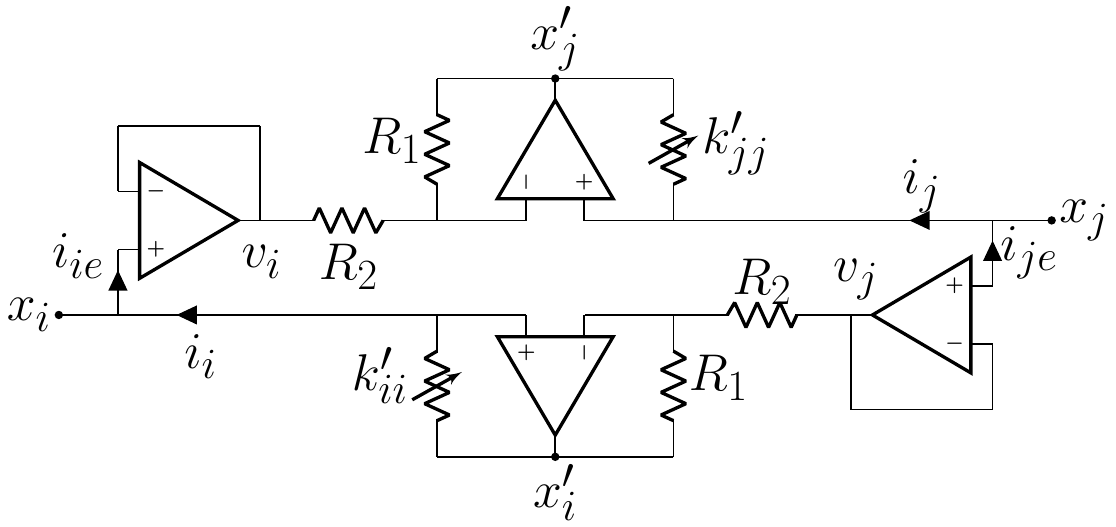} 
\caption{Negative Resistance Circuit with buffers $i_{ie}=i_{je}=0$}
\label{fig:NegRes} 
\end{figure}

Consider the negative resistance between nodes $i$ and $j$ with a conductance of $k_{ij}$, where the voltage at nodes $i$ and $j$ are denoted by $x_i$ and $x_j$. 
Instead of directly connecting the two nodes with the resistor $k_{ij}$, this design creates two nodes $i'$ and $j'$ such that node $i$ is connected to node $i'$ by the resistor $k_{ii}'$ and node $j$ is connected to node $j'$ by the resistor $k_{jj}'$. 
The conductance of the new resistors and the voltage of the two new nodes are designed to be as follows:
\begin{align}
k_{ij} &= k_{ii}' = k_{jj}'\\
x_i'&= x_j + 2( x_i - x_j )\\
x_j'&= x_i + 2( x_j - x_i )
\end{align}
These equations guarantee that if $x$ amount of current is to flow out of node $i$ into node $j$ in the case of normal resistor $k_{ij}$, the same $x$ amount of current will flow into node $i$ from node $i'$ and the same $x$ amount of current will flow out of node $j$ into node $j'$. 
The current flowing out of nodes $i$ and $j$ in the negative resistance design is as follows:
\begin{align}
i_i &= k_{ii}' ( x_i' - x_i ) = -k_{ij} ( x_j - x_i ) \\
i_j &= k_{ii}' ( x_j' - x_j ) = -k_{ij} ( x_i - x_j )
\end{align}
To create the two nodes $i'$ and $j'$, the non-inverting operational amplifier configuration is used as shown in Figure \ref{fig:NegRes} with negative feedback loop. 
The two resistors $R_1$ and $R_2$ are chosen to have equal conductance to have a gain of 2. 
The circuit in Figure \ref{fig:NegRes} shows the amplifier circuits together with the two resistors $k_{ii}'$ and $k_{jj}'$. 
Buffer circuits are added to stop unaccounted currents from being drawn from / supplied to the nodes where computed currents will be drawn from / supplied to.

The governing differential equation of the proposed negative resistance design is derived from the transfer functions for the individual OAs. 
For the upper part of the circuit, the transfer functions in the $s$-domain for the buffer OA and the gain-2 OA are
\begin{align}
v_i(s)&=[x_i(s)-v_i(s)]L(s) \label{equ:v_i(s)}\\
x_j'(s) &= \left[x_j(s)-\frac{x_j'(s)+v_i(s)}{2}\right]L(s) \label{equ:x_j'(s)}
\end{align}
where $L(s)$ is the OA open-loop transfer function. 
The single-pole OA transfer function is defined as follows
\begin{equation}
L(s)=\frac{L_0}{1+s/\omega_0} = \frac{1}{1/L_0+s/(L_0\omega_0)}\approx \frac{L_0\omega_0}{s} \label{equ:L(s)}
\end{equation} 
where $L_0$ is the DC open-loop gain and $\omega_0$ is the 3-dB bandwidth \cite{OA_book}.
Defining $\tau=1/(L_0\omega_0)$ and combining Equations \eqref{equ:v_i(s)}, \eqref{equ:x_j'(s)} and \eqref{equ:L(s)} result in the following
\begin{equation}
2\tau s x_j' = 2x_j-x_j'-\frac{x_i}{1+\tau s}
\end{equation}
The equation is simplified to the following form in the $s$-domain 
\begin{equation}
(2\tau^2s^2+3\tau s+1)x_j'=2x_j-x_i+2\tau sx_j
\end{equation}
Transforming the equation into the time domain gives the governing second-order differential equation as follows
\begin{equation}
2\tau^2\frac{d^2x_j'}{dt^2}+3\tau \frac{dx_j'}{dt}+x_j'=2x_j-x_i+2\tau \frac{dx_j}{dt} \label{equ:R_neg_DE}
\end{equation}

Another, further simplified differential equation that can be used to estimate the lower bound of the settling time is derived by assuming that the buffer OA is an ideal OA with infinite bandwidth. 
This assumption ignores the time delay in the buffer OA stage, where the transfer function in Equation \eqref{equ:v_i(s)} is modified to $v_i=x_i$. 
Following the same derivation steps, the resulting first-order governing differential equation for the lower bound of the settling time is as follows
\begin{equation}
2\tau \frac{dx_j'}{dt}+x_j'=2x_j-x_i
\label{equ:R_neg_DE_bound}
\end{equation}


\section{Preliminary Design: $n$-unknowns}\label{secDes1}

\begin{figure}[!t]  
\centering 
\includegraphics[page=4,width=0.5\textwidth]{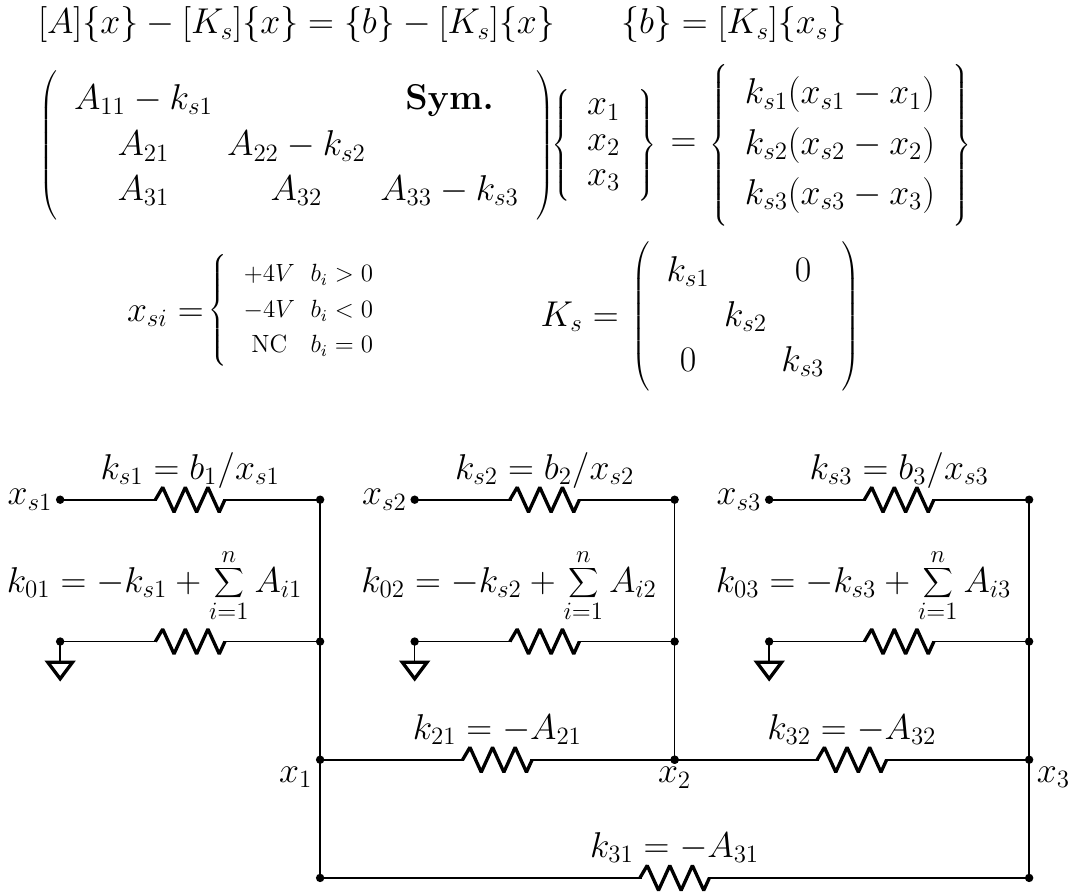} 
\caption{Resistive network for a 3x3 system $(A-K_s)x=b-K_sx$. The current sources are replaced with voltage sources and resistors.}
\label{fig:A-Ksx=b-Ksx} 
\end{figure}

In this design, the given symmetric system of equations $Ax=b$ is mapped to an equivalent system of resistors, where the unknown vector $x$ represents the voltages at the nodes. 
To simplify the system and avoid the incorporation of active dynamic components, the right-hand-side vector $b$ is not directly assigned as the supplied current. 
Instead, a diagonal matrix $K_s$ with strictly nonnegative coefficients is introduced. 
This matrix represents the conductance of the resistors that connect the supply nodes to the nodes corresponding to the unknowns. 
These resistors have conductance values denoted as $k_{si}$. 
The introduction of $K_s$ modifies the original system by subtracting this matrix from both sides of the equation, resulting in the following form:
\begin{equation}
(A-K_s)x = b-K_sx
\end{equation}
The vector $b$ is redefined as $b=K_s x_s$, where $x_s$ represents the voltage vector of the supply nodes. These supply voltages, $x_{si}$, are connected to their respective unknown node voltages $x_i$ through resistors with conductance $k_{si}$. The representation of a given symmetric system of equations by a resistive network using the preliminary design is illustrated in Figure \ref{fig:A-Ksx=b-Ksx}.

\subsection{Right-Hand-Side: $(b-K_s x)$}\label{subsecDes1RHS}

The right-hand side of the new system represents the current flowing from the supply nodes, which can be adjusted by controlling either the voltage of the supply nodes or the conductance of the associated resistors. In this design, the voltages of the supply nodes are kept constant and the vector $b$ is mapped by varying the conductance of the supply resistors. This choice eliminates the need for a Digital-to-Analog Converter (DAC) and simplifies the implementation, as the control of variable resistors is also needed for the left-hand-side.

\begin{figure}[!t]  \centering 
\includegraphics[page=5,width=0.3\textwidth]{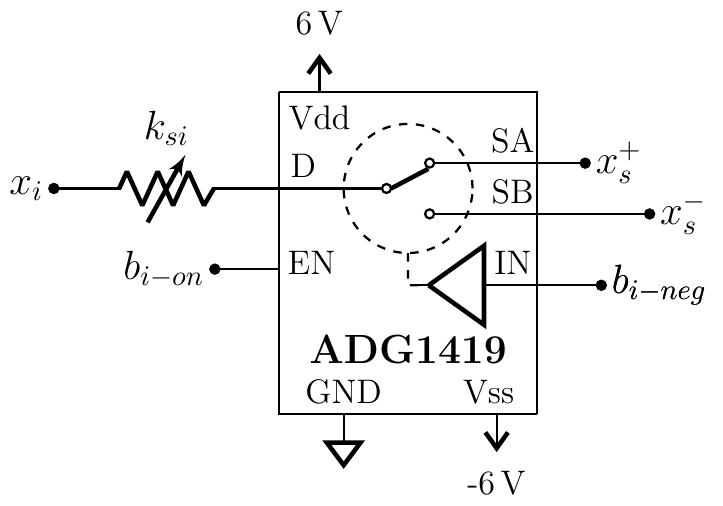} 
\caption{Right-Hand-Side Circuit for each node. Each node $i$ is connected to a supply resistor of conductance $k_{si}$ that is connected to either $x_s^+$ in case $b_i$ is positive or $x_s^-$ in case $b_i$ is negative. In the case of $b_i=0$, the switch is not enabled.}
\label{fig:RHS-Des1} 
\end{figure}

The supply nodes are selected to be either $x_s^+=4V$ or $x_s^-=-4V$. Each unknown node $i$ with voltage $x_i$ is connected to the Drain terminal (D) of a digitally controlled analog switch through a variable resistor with conductance $k_{si}$. The operation of the switch is controlled by two digital signals:

\begin{enumerate}
    \item \textbf{Enable Terminal (EN):} Connected to the digital bit $b_{i-on}$. If the corresponding element of the vector $b$ is zero, $b_{i-on}$ is set to digital zero, causing the switch to remain open and isolating the node $i$ from any supply node. 
    \item \textbf{Control Terminal (IN):} Connected to digital bit $b_{i-neg}$. This bit determines whether the node is connected to $x_s^+$ (positive supply) via the Source A terminal (SA) for positive elements of $b$ or to $x_s^-$ (negative supply) via the Source B terminal (SB) for negative elements of $b$.
\end{enumerate}

\noindent The conductance of each supply resistor, $k_{si}$, is computed as follows:
\begin{equation}
k_{si} = b_i / x_{si} = |0.25 b_i|
\label{equ:k_si}
\end{equation}
This computation can be done digitally by two-bit shift or by directly connecting the bits to a digital potentiometer with two-bit shift. An example of the right-hand-side circuit for node $i$ is shown in Figure \ref{fig:RHS-Des1}.

\subsection{Left-Hand-Side: $(A-K_s)x$}\label{subsecDes1LHS}

\begin{figure}[!t]  \centering 
\includegraphics[page=6,width=0.5\textwidth]{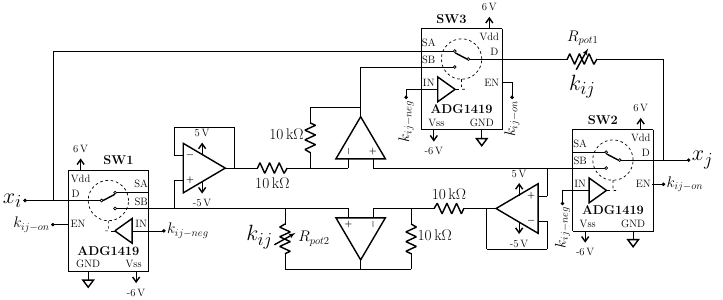}  
\caption{Element Circuit. The three switches are enabled when the required conductance $k_{ij}$ between node $i$ and $j$ is not zero. If $k_{ij}$ is positive the active circuit becomes only the resistor $R_{pot1}$ connecting nodes $i$ and $j$. If $k_{ij}$ is negative the active circuit becomes the negative resistance circuit between nodes $i$ and $j$}
\label{fig:ElemCirc} 
\end{figure}

The left-hand side of the new system represents the conductance matrix of the resistive network, where each resistive element in the circuit can exhibit either positive or negative resistance. 
To ensure the generality of the circuit and its ability to represent any symmetric system of equations, this design employs digitally controlled analog switches to toggle between a normal variable resistor (positive resistance) and a negative resistance circuit for each element of the system, as shown in Figure \ref{fig:ElemCirc}.

For each resistive element in the matrix $(A-K_s)$, the resistance values of the variable resistors $R_{pot1}$ and $R_{pot2}$ are set to be the absolute value of the inverse of the corresponding conductance calculated by Equations \eqref{equ:kij} and \eqref{equ:ki0}. Two additional control bits are associated with each resistive element:

\begin{enumerate}
    \item \textbf{Activation Bit} ($k_{ij-on}$): This bit determines whether the resistive element is active (i.e., has nonzero conductance). It is connected to the Enable (EN) terminals of the three switches within the element circuit.
    \begin{itemize}
        \item when $k_{ij-on}=0$, the switches remain open, disconnecting nodes $i$ and $j$, effectively removing the element from the circuit.
        \item when $k_{ij-on}=1$, the switches close, connecting the Drain terminal (D) to either the Source terminal A (SA) or the Source terminal B (SB), depending on the second control bit. 
    \end{itemize}

    \item \textbf{Polarity Bit} ($k_{ij-neg}$): This bit determines whether the element represents a positive or negative resistance. It is connected to the control terminals (IN) of the three switches within the element circuit.
    \begin{itemize}
        \item when $k_{ij-neg}=0$, the switches configure the circuit for positive resistance. The Drain terminal (D) of the switches are connected to the Source terminal A (SA), establishing a connection between nodes $i$ and $j$ only through the variable resistor $R_{pot1}$. In this configuration, switch SW3 closes to activate the positive resistance path, while switches SW1 and SW2 remain open, isolating the negative resistance circuit. 
        \item when $k_{ij-neg}=1$, the switches configure the circuit for negative resistance. The Drain terminal (D) of the switches are connected to the Source terminal B (SB), engaging the negative resistance circuit between nodes $i$ and $j$.
    \end{itemize}
\end{enumerate}

\begin{figure}[!t]  \centering 
\includegraphics[page=7,width=0.5\textwidth]{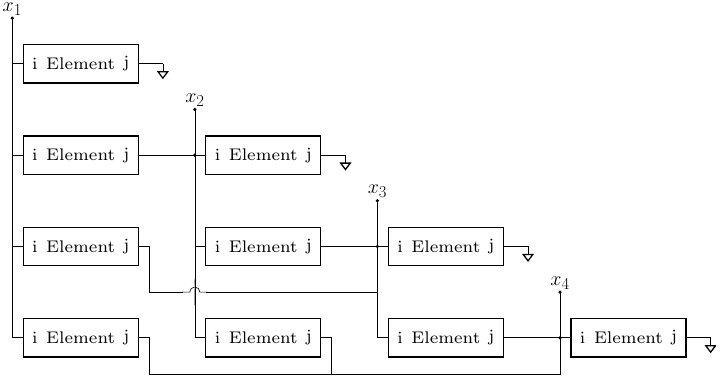} 
\caption{The assembly of the Left-Hand-Side Circuit. Each element has digital inputs representing $k_{ij-on}$, $k_{ij-neg}$, and $k_{ij}$}
\label{fig:LHS-Des1} 
\end{figure}

\noindent The assembly of the left-hand-side of the design closely resembles a lower triangular matrix, reflecting the symmetry of the system. Each element in the off-diagonal of the linear system corresponds directly to an element circuit connecting the same pair of nodes. For instance, the element in row 2, column 3 of the matrix and the layout is represented by an element circuit between the unknowns $x_2$ and $x_3$. The diagonal elements in this layout represent the elements between the unknown $x_i$ and the ground. This structured layout helps visualizing the relation between the circuit elements and the matrix elements. An example of the assembly of the left-hand-side of this design for a 4x4 system is shown in Figure \ref{fig:LHS-Des1}.

\subsection{System Testing}\label{subsecDes1Tests}

To evaluate the performance of the system, random symmetric matrices of varying sizes were generated using MATLAB R2022a \cite{matlab} and then simulated in LTspice v.17.1.8 \cite{ltspice}. The simulations were transient analysis simulations where the supply voltage was modeled as a step function that transitions from zero to $x_s^+$ or $x_s^-$ at 1 $\mu s$. This setup was used to analyze the system settling time and the accuracy of the solution for the unknown vector $x$. The observed performance of the preliminary design across different subdomains of the linear system of equations $Ax=b$, is summarized as follows:

\begin{enumerate}
    \item \textbf{Unsymmetric Systems:}
    
    These systems cannot be represented by the circuit. A transformation is required to convert such systems into Symmetric Positive Definite (SPD) systems. One potential transformation, though computationally expensive, is $A^TAx=A^Tb$.\\
    
    \item \textbf{Symmetric Non-Positive-Definite Systems:}

    \begin{figure}[!t]  \centering
    \includegraphics[page=8,width=0.5\textwidth]{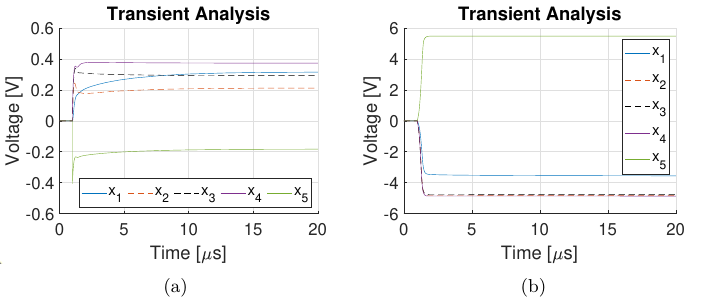}  
    \caption{Transient Analysis of two 5x5 systems using the preliminary design. Plot (a) shows a stable solution for a Positive Definite matrix ($Ax=b$). Plot (b) shows unstable solution for a Negative Definite matrix ($-Ax=-b$). The theoretical solution of both systems is $x = [0.32\ \ 0.21\ \ 0.29\ \ 0.37\ \ -0.18]^T$.}
    \label{fig:TransOpamp} 
    \end{figure}

    While these systems can be represented by the circuit, the system does not converge to the expected solution. 
    In such cases, the voltage at the output node of at least one operational amplifier (OA) in the negative resistance circuits reaches the maximum or minimum output voltage of the OA. 
    This saturation indicates a positive feedback loop that results in circuit instability. This instability observation can be explained by using the analogy between resistive networks and spring networks as presented in \cite{MacNeal}. 
    In mechanical systems, a non-positive eigenvalue in the stiffness matrix (analogous to the conductance matrix) signifies instability, where small perturbations in displacement (analogous to voltage) result in unbounded response.

    A test case for a 5x5 system is illustrated in Figure \ref{fig:TransOpamp}. 
    The system is symmetric positive definite and is solved in two scenarios. 
    The first plot presents the solution for the original system, while the second plot shows the results of the negative definite system obtained by multiplying both the conductance matrix $A$ and the vector $b$ of the original system by -1. 
    In theory, the expected solution for both systems should be identical. The solution vector for the systems is $x = [0.32\ \ 0.21\ \ 0.29\ \ 0.37\ \ -0.18]^T$. The Negative Definite system exhibits instability, whereas the Positive Definite system demonstrates a stable solution with a maximum error of less than 0.01\% relative to the expected solution. \\

    \item \textbf{Symmetric Positive-Definite Systems:}
    
    These systems can be represented by the circuit, and the circuit converges to the expected solution. 
    The accuracy of the solution depends on the specifications of the opamps and variable resistors used. 
    For systems where at least one conductance value is negative, the active circuit of the system includes opamps that require a settling time before stabilizing. 
    
    To evaluate convergence time and error in this domain, random symmetric systems of varying sizes were generated and simulated. 
    The conductance matrices, created using MATLAB's "sprandsym" function, are full matrices with eigenvalues ranging between and including 10 $\mu S$ and 1000 $\mu S$. 
    For each generated system, the unknowns $x_i$ were randomly assigned within $[-0.5V, 0.5V]$, and the corresponding $b$ vector was computed by $b=Ax$. 
    An equivalent circuit for each generated system was modeled and simulated in LTspice v.17.1.8 software using the AD712 model for the opamps through two analyses:

    \begin{figure}[!t]  \centering
        \includegraphics[page=9,width=0.5\textwidth]{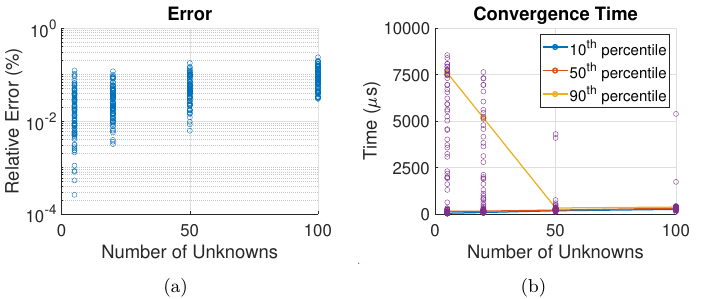}  
        \caption{Analysis of 400 random dense systems (eigenvalues: [10 1000] $\mu S$) using the preliminary design. Plot (a) shows the maximum error between the converged solution and the theoretical values of each system. Plot (b) shows the maximum convergence time for each system. }
        \label{fig:ComplexOpamp} 
    \end{figure}

    \begin{itemize}
        \item \textbf{Operating Point Analysis:} The supply nodes had constant voltage of $x_s^+$ and $x_s^-$. 
        This analysis provided the expected converged solution for the circuit. 
        The maximum error in the converged solution of each generated system is shown in Figure \ref{fig:ComplexOpamp}. 
        In all simulated systems, the expected errors from using this design is less than 0.3$\%$.
        \item \textbf{Transient Analysis:} The supply voltage was modeled as a step function transitioning to $x_s^+$ and $x_s^-$ at 1 $\mu s$. 
        Convergence time was defined as the time instant beyond which the voltages are always within 1$\%$ of the operating point solution. 
        The results from using the element circuit shown in Figure \ref{fig:ElemCirc} are presented in Figure \ref{fig:ComplexOpamp} and show median convergence times ranging from 144 $\mu s$ for 5x5 systems to 307 $\mu s$ for 100x100 systems. 
        However, the $90^{th}$ percentile times range from 7600 $\mu s$ to 370 $\mu s$, reflecting large variability (an order of magnitude difference). 
        This variability, a result of the specific choice of the element circuit, complicates determining when to read reliably the solution in practical applications and highlights the need for a better design in this domain.
    \end{itemize}
    
    \item \textbf{Positive Weighted Undirected Graph Systems:}
    
    A subset of symmetric positive definite systems, these systems can be decomposed into the sum of a Laplacian matrix and a diagonal matrix with strictly positive values. 
    The active circuit for such systems consists solely of positive resistors, enabling a purely resistive network. 
    As a result, the solution is achieved at the theoretical maximum speed determined by the current flow in the network ($O(1)$) regardless of the problem size. 
    The accuracy in this case depends solely on the specifications of the variable resistors.

\end{enumerate}


\section{Proposed Design: $2n$-unknowns}\label{secDes2}

The primary motivation for this design is to reduce the number of components in the circuit, increase the range of problems that can be represented using strictly positive resistance, and reduce the settling time of the circuit. 
The design achieves this by transforming the original $n\times n$ system of equations $Ax=b$ into a new $2n\times 2n$ system that ensures that most resistive elements are positive, without introducing significant additional digital computations. The transformed system $Ku=f$ is expressed as:
\begin{equation}
\begin{bmatrix}
K_A & K_B\\
K_B & K_A
\end{bmatrix}
\begin{Bmatrix}
x \\ -x
\end{Bmatrix}
=
\begin{Bmatrix}
b - K_s x \\ -b-K_s(-x)
\end{Bmatrix}
\label{equ:Des2sys}
\end{equation}
 where $K_A$ and $K_B$ are $n$x$n$ sub-matrices defined as:
\begin{align}
K_A &= D+0.5(A-|A|)-K_s\\
K_B &= D-0.5(A+|A|)
\end{align}
where $D$ is a diagonal matrix with strictly non-negative elements. 
This transformation ensures that the off-diagonal terms of the sub-matrices $K_A$ and $K_B$ are negative, corresponding to positive resistive elements in the circuit. 
Importantly, no additional computations are required for these values, as they directly correspond to the off-diagonal terms of the original $A$ matrix, although assigned between different nodes. 
For positive resistive element between nodes $i$ and $j$ in the original system, the transformed system retains the same positive resistance between these nodes. 
Additionally, it introduces another positive resistive element with identical conductance between nodes $n+i$ and $n+j$. 
For negative resistive elements between nodes $i$ and $j$ in the original system, the transformed system replaces the negative resistance with two positive resistive elements of the same conductance. 
One resistor connects nodes $i$ and $n+j$, while the other connects nodes $j$ and $n+i$.
Thus, in the transformed $2n\times2n$ system, the only off-diagonal terms that can possibly have a positive value (i.e. negative resistance), are the $n$ diagonal elements of $K_B$. This is a significant reduction from the possibly $n(n+1)/2$ negative resistances in the preliminary design shown in Figure \ref{fig:LHS-Des1}. 

\subsection{Choice of $D$-Matrix}\label{subsec:D-mat}

The choice of matrix $D$ plays a pivotal role in determining the domain of symmetric linear systems that will be Positive Definite (PD) in the transformed system and, therefore, solvable. 
In addition, it influences both the speed and the accuracy of the solution. 
A comparable, though less practical, transformation was proposed for support trees \cite{GrembanPHD}. 
In that approach, $D$ is selected as the diagonal of the original matrix $A$, and $K_s$ is set to the zero matrix. 
However, such approach does not guarantee that a positive definite system will remain positive definite after transformation, thereby reducing the range of problems that can be addressed. 

\subsubsection{Positive Definiteness Condition}

To ensure the stability of the system, it is essential to analyze the conditions under which the transformed system remains Positive Definite (PD), as this is a key criterion for stability. 
To find the eigenvalues of the transformed matrix $K$, we define an orthogonal matrix $Q$ that diagonalizes the block matrix $K$ as follows
\begin{align}
Q &= \frac{1}{\sqrt{2}} \begin{bmatrix}I & I \\ -I & I \end{bmatrix} \\
QKQ^T &= \begin{bmatrix}K_A+K_B & 0 \\ 0 & K_A-K_B \end{bmatrix} \label{equ:QKQ}
\end{align}
The eigenvalues $\lambda$ of the transformed system, for any choice of $D$, are computed as follows:
\begin{equation}
\det(K-\lambda I) = \det(K_{AB}-\lambda I_n)\ \det(K_A-K_B-\lambda I_n) 
\end{equation}
where $K_{AB}=K_A+K_B$. 
Thus, positive definiteness of the system is achieved if both matrices $(K_A-K_B)$ and $K_{AB}$ are PD. 
These matrices can be expressed and simplified as follows:
\begin{align}
K_A - K_B &= A - K_s \label{equ:KA-KB}  \\
K_{AB} =K_A + K_B &= 2D - |A| - K_s
\end{align}

The condition expressed in Equation \eqref{equ:KA-KB} requires that the original system $(A-K_s)$ be PD. 
To preserve the positive definiteness of the transformed system, the matrix $D$ is chosen to ensure that $(K_A+K_B)$ is diagonally dominant, making it PD. 
Since $(K_A+K_B)$ can be decomposed as the sum of a diagonal matrix $D$ with strictly positive elements and a matrix $(-|A|-K_s)$ with strictly non-positive elements, achieving diagonal dominance requires that the sum of each column of $(K_A+K_B)$ be greater than zero. 
This requirement leads to the following condition for each element of the matrix $D$:
\begin{equation}
    D_{ii} > \frac{1}{2} \left[ (K_{s})_{ii} +  \sum_{j=1}^{n} |A_{ji}| \right]
\label{equ:D_condition}
\end{equation}

\subsubsection{Scaling}

To optimize the design for the choice of $D$, the effect of the $D$ matrix on both the accuracy and speed of the expected solution was studied. 
In this study, $D$ was selected as a scaled identity matrix, with the scaling determined as the maximum value of the sums of the columns of the absolute value of the conductance matrix $A$ as follows: 
\begin{equation}
    D = \beta \max_i \left( \sum_{j=1}^{n} |A_{ji}| \right) I_n
\end{equation}
where $\beta$ is a scaling factor with $\beta \geq 0.5$ to satisfy the condition in Equation \eqref{equ:D_condition}.

\begin{figure}[!t]  \centering
    \includegraphics[page=10,width=0.5\textwidth]{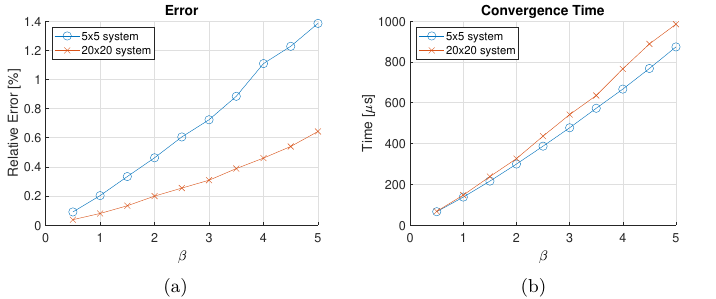}  
    \caption{Analysis of two random systems using the proposed design for different values of $D$ matrix scaling $\beta$. Plot (a) shows the maximum error between the converged solution and the theoretical values of each system. Plot (b) shows the maximum convergence time for each system. }
    \label{fig:ParametricBeta} 
\end{figure}

Random symmetric positive definite matrices, not diagonally dominant, with variable sizes were generated and simulated for various values of $\beta$. The results consistently showed that reducing $\beta$ significantly decreased both the error and the convergence time, as illustrated for two example systems in Figure \ref{fig:ParametricBeta}. Thus, minimizing $D$ while ensuring that the system remains positive definite is favorable for accuracy and performance.

\subsubsection{Proposed $D$-Matrix}\label{subsubsec:Prposed-D-Mat}

The selection of $D$ was further refined using an analogy between resistive networks and 1D spring systems. A conductance (stiffness) matrix $A$ with a zero determinant (at least one zero eigenvalue) corresponds to a spring system without supports, making it unstable. In a resistive network (1D Spring System), this instability occurs when the sums of the columns in the matrix are zero. Stability can be achieved by ensuring that at least one column sum is positive, akin to providing a single support in a mechanical system.

To stabilize the system, the diagonals of $D$ were chosen as:
\begin{equation}
    D_{ii} = 
    \begin{cases}
        (K_{s})_{ii} +  \frac{1}{2} \left[ \sum_{j=1}^{n} |A_{ji}| \right] & \text{for}\ i=1\\\\
        \frac{1}{2} (K_{s})_{ii} + \frac{1}{2} \left[ \sum_{j=1}^{n} |A_{ji}| \right] & \text{for}\ i\neq 1
    \end{cases}
    \label{equ:Dii}
\end{equation}
This choice ensures that the sum of each column in the matrix $(K_A+K_B)$, except for the first column, is zero. It also simplifies the design by ensuring that only nodes 1 and $n+1$ are connected to the ground. The conductance between these nodes and the ground is $k_{s1}$, which is the conductance of the resistors connecting these nodes to the supply node $x_s$.

The choices for the system transformation in the proposed design do not introduce additional digital computations beyond those required in the preliminary design. The resistive network for the proposed design is constructed with conductance values assigned as:

\begin{equation}
k_{ij} =
\begin{cases}
k_{s1} & \text{for} \>\> i=0 ,  \\ & \text{and} \>\> j=1, \> n+1 \\
-A_{ij} & \text{for} \>\>  j <i ,\> A_{ij}<0 \\ & \text{and} \>\> i =1,\cdots,n \\
A_{ij} & \text{for} \>\>  j < i ,\> A_{ij}>0 \\ & \text{and} \>\> i=1,\cdots,n \\
A_{ii} - \frac{1}{2} \left(  k_{si} +  \sum\limits_{c=1}^{n} |A_{ci}|  \right) & \text{for} \>\> j =n+i \\ & \text{and} \>\> i=1,\cdots,n \\
0 & \text{for all other elements}  \\
\end{cases}
\end{equation}

\noindent This approach optimizes the stability, accuracy, and convergence speed of the system while minimizing unnecessary complexity in the design.

\begin{figure}[!t]  \centering
    \includegraphics[page=11,width=0.5\textwidth]{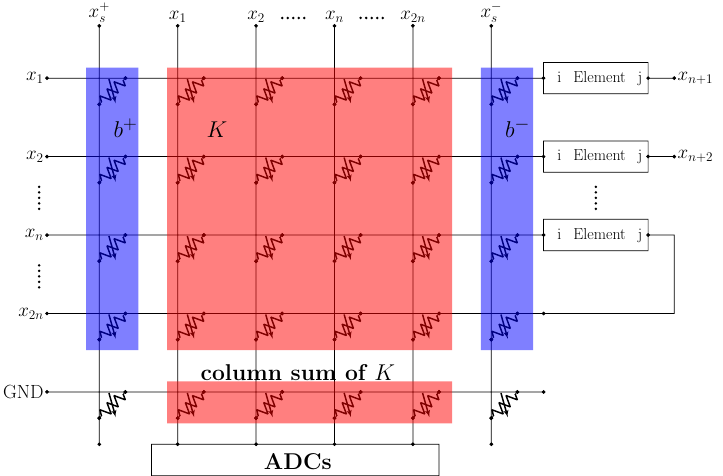}  
    \caption{
    Proposed design in cross-point architecture. Entries of the transformed $2n\times2n$ matrix $K$ are directly assigned to the memristor array. 
    The word lines (horizontal) and the bit lines (vertical) corresponding to the same entry of the unknown vector $x$ are connected. 
    Two bit lines of memristors are added to represent the positive and negative entries of the right-hand side vector $b$. 
    One word line is added to allow assigning conductance between each node and the ground. 
    Peripheral circuitry representing the element circuit is also added between nodes $x_i$ and $x_{n+i}$ to account for any negative resistances corresponding to positive diagonals of $K_B$.  
    }
    \label{fig:crosspoint} 
\end{figure}

\subsection{Cross-point Architecture}

A notable advantage of the proposed design is its compatibility with cross-point architecture, as shown in Figure \ref{fig:crosspoint}, the same architecture that is widely used for Matrix Vector Multiplication (MVM). 
In this setup, the rows and columns of the array represent the voltages of the unknown vector $x$ (and $-x$), with each row directly connected to the corresponding column. 
Two additional columns are introduced for the supply nodes $x_s^+$ and $x_s^-$, and an additional row is included for the ground node.

Elements of the transformed matrix $K$ correspond to the conductance values between the various nodes in the unknown vector $x$. 
Theoretically, only half of the elements in the cross-point array are needed to represent the system due to symmetry. However, to maintain the matrix structure in the array for MVM operations, the off-diagonal terms of the submatrices $K_A$ and $K_B$ are divided by two. 
These values are then assigned symmetrically to the elements $k_{ij}$ and $k_{ji}$, effectively splitting the original resistors into two parallel resistors. 
In practical memristor-based implementations, double precision of the conductance values can be achieved by employing read–verify feedback programming across multiple arrays (typically no more than five) \cite{song2024programming}.

The diagonal elements of the transformed matrix $K$, whether assigned within the array or not, do not influence the design, as both ends of these resistors are connected to the same node, rendering them effectively nonexistent. 
Furthermore, the diagonals of the submatrix $K_B$ are deliberately switched off in the array, as these elements can represent either positive or negative resistance. 
External element circuits are used to connect each node $i$ to node $n+i$, enabling flexible resistance representation while preserving the structure of the array for MVM operations.

The elements of the supply matrix $K_s$ are assigned to the conductance values in the two additional columns, based on the sign of the corresponding element in the right-hand-side vector $b$. 
Positive elements $b^+$ are connected to $x_s^+$ whereas negative elements $b^-$ are connected to $x_s^-$. 
The conductance values are calculated using Equation \eqref{equ:k_si}. 
The additional row in the system includes the conductance values calculated as the sums of the columns of the transformed matrix $K$. 

Once the circuit reaches steady state, the solution vector is obtained by reading the $n$ node voltages corresponding to the unknowns using ADCs connected to those nodes.
For large arrays, the number of ADCs can be reduced by employing multiplexers along with the ADCs. 
Instead of reading all $n$ unknowns simultaneously, the solution voltages can be sampled serially in groups of $m$ nodes, where $m$ corresponds to the number of available ADCs. 

To enable MVM operation on the proposed architecture, programmable switches can be introduced to reconfigure the peripheral circuitry between the element circuitry and the current sensing configuration required for MVM.
Specifically, a set of switches selects between the peripheral element circuits and the operational amplifiers used for current sensing, while an additional set of switches on the word lines (WLs) temporarily disconnects the direct WL–bit line (BL) connections during MVM mode.
The input vector for the MVM operation is applied by appropriately programming the $b^+$ and $b^-$ conductances using two-bit shifts, assuming symmetric supply voltages of $+4\mathrm{V}$ and $-4\mathrm{V}$.

\subsection{Circuit Dynamics}\label{subsecCircDyn}

A formal representation of the proposed design can be established by decomposing the transformed matrix $K$ into two matrices, $P$ and $N$, which correspond to the conductance matrices of the positive and negative resistance elements, respectively. 
The transformed system of equations $Ku=f$ can then be expressed as
\begin{equation}
    (P-N) u =f
\end{equation}

\subsubsection{Governing Differential Equation}
\hspace*{\fill} \par
The proposed design guarantees that there is no more than one negative resistance circuit at each node $j$ in the system connected to node $n+j$. 
Let the diagonal matrix $R$ be the matrix of conductance values of the negative resistance circuits connected at each node. 
This matrix can be defined as the positive diagonals of the matrix $K_B$. 
The matrices $R$, $K_N$, $N$, and $P$ are computed as follows:
\begin{align}
R &= \frac{1}{2}(K_B+\lvert K_B\rvert) \\
K_N &= \begin{bmatrix} R & 0 \\ 0 & R \end{bmatrix} \\
N &= \begin{bmatrix} R & -R \\ -R & R \end{bmatrix} \\
P &= K+N
\end{align}

Applying Equation \eqref{equ:R_neg_DE} for the transformed system and using the voltage relationship $u_j=-u_{n+j}$, the governing differential equation in vector form can be written as
\begin{equation}
2\tau^2 \frac{d^2u'}{dt^2}  +3\tau \frac{du'}{dt}  -2\tau \frac{du}{dt} = 3u - u'  \label{equ:DE_of_u'}
\end{equation}
At any time $t$, the proposed design solves the resistive network that is composed from the positive conductance matrix $P$. 
The current supplied to the network is the summation of the current represented by the $f$-vector and the current supplied from the negative resistance circuit.
The system in matrix form can be expressed as:
\begin{equation}
Pu=f+K_N(u'-u) \label{equ:DE_at_t}
\end{equation}
Equation \eqref{equ:DE_at_t} can be rearranged as follows 
\begin{equation}
K_Nu' = Mu - f 
\end{equation}
where $M = (P+K_N)$. 
Taking the first and second derivatives of the equation results in the following:
\begin{align}
K_N\frac{du'}{dt} &= M\frac{du}{dt} \label{equ:du'dt}\\
K_N\frac{d^2u'}{dt^2} &= M\frac{d^2u}{dt^2} \label{equ:d2u'dt2}
\end{align}
Using the relationship $Nu=2K_Nu$ that arises from the structure of the vector $u$, Equation \eqref{equ:DE_at_t} can be rewritten as:
\begin{equation}
Ku-f = -K_N (3u - u') \label{equ:3u-u'}
\end{equation}
Substituting Equations \eqref{equ:du'dt}, \eqref{equ:d2u'dt2}, and \eqref{equ:3u-u'} in Equation \eqref{equ:DE_of_u'} gives:
\begin{equation}    
2\tau^2 M \frac{d^2u}{dt^2}  +3\tau M\frac{du}{dt} -2\tau K_N\frac{du}{dt}  = -(Ku-f)
\label{equ:DE_2nd_order}
\end{equation}
To solve the second-order differential equation using state space, a new vector $w$ is defined as follows:
\begin{align}
w &=\tau \frac{du}{dt} \\
\frac{dw}{dt} &=\tau \frac{d^2u}{dt^2}
\end{align}
Thus, the differential Equation \eqref{equ:DE_2nd_order} can be written as follows:
\begin{equation}
 -2\tau\frac{dw}{dt} = M^{-1}(Ku-f) + (3I-2M^{-1}K_N) w
\end{equation}
The matrix form of the complete differential equations:
\begin{equation}
\frac{d}{dt} \begin{bmatrix}
u \\ w
\end{bmatrix} = \frac{-1}{2\tau} \left( S \begin{bmatrix}
u \\ w
\end{bmatrix} \\ - \begin{bmatrix}
0 \\ M^{-1}f
\end{bmatrix} \right)
\end{equation}
where
\begin{equation}
S = \begin{bmatrix}
 0 & -2I \\
 M^{-1}K & 3I-2M^{-1}K_N
\end{bmatrix}
\end{equation}
Using the finite difference, the governing equations become
\begin{equation}
\begin{bmatrix}
u(t+\Delta t) \\ w (t+\Delta t)
\end{bmatrix} = \frac{\Delta t}{2\tau} 
\begin{bmatrix} 0 \\ M^{-1}f \end{bmatrix} 
+\left( I - \frac{\Delta t}{2\tau}  S \right) 
\begin{bmatrix} u(t) \\ w(t) \end{bmatrix}
\end{equation}

\begin{equation}
\begin{bmatrix} u(0) \\ w (0) \end{bmatrix} = 
\begin{bmatrix} M^{-1}f \\ 0 \end{bmatrix}
\end{equation}

\begin{figure}[!t]  \centering
    \includegraphics[page=12,width=0.5\textwidth]{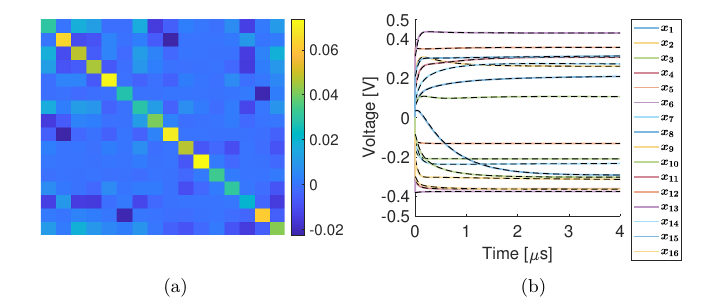}  
    \caption{Differential Equation Validation. (a) Matrix $A$ of a random 16x16 linear system of equations that is composed of a mix of Diagonally Dominant (DD) rows and non-DD rows. (b) Transient solution of the system using LTSpice simulation (color lines) and the finite difference algorithm of the derived differential equation (black dashed lines) }
    \label{fig:DiffEqVal} 
\end{figure}

To validate the derived differential equation governing the circuit dynamics, 1000 random systems were generated and simulated in transient analysis using LTspice v.17.1.8. 
In these simulations, the OAs are modeled as single-pole models with a 10 MHz gain-bandwidth product. 
Across all tested systems, the transient response predicted by the differential equation matches within 0.2\% the transient output voltages produced by the LTspice simulation. 
A representative example for a 16x16 system of equations containing a mixture of Diagonally Dominant columns (DD) and non-DD ones is shown in Figure \ref{fig:DiffEqVal}.

\subsubsection{Pole Analysis}
\hspace*{\fill} \par
To characterize the dynamic behavior of the circuit, the poles of the governing differential equation are analyzed. 
The dynamics of the full $2n$ system $K$ decouple into two independent $n$ subsystems, which arise from the diagonalization of the transformed matrix $K$ as shown in Equation \eqref{equ:QKQ}. 
The first subsystem is associated with the matrix $K_{AB}$, whose eigenvectors lie in the vector space $\begin{bmatrix}I_n & I_n\end{bmatrix}^T$. 
As the unknown vector $u$ is orthogonal to this vector space, the convergence and stability of the solution process is unaffected by the dynamics of this subsystem. 
Nevertheless, it is essential that the $K_{AB}$ subsystem is not divergent, ensuring that perturbations do not grow and destabilize the circuit. 
The second subsystem corresponds to the system matrix $A$. 
This system governs the actual convergence behavior of the solver circuit.

The oscillation and stability conditions for the $K_{AB}$ system, given in the Appendix (Equations \eqref{equ:K_AB_osc} and \eqref{equ:K_AB_stab}), depend on the eigenvalues of the matrices $R$ and $K_{AB}^{-1}$. 
Since the matrix $R$ is a nonnegative diagonal matrix and the matrix $K_{AB}$ is diagonally dominant, both matrices have nonnegative eigenvalues. 
Consequently, the conditions in Equations \eqref{equ:K_AB_osc} and \eqref{equ:K_AB_stab} confirm that perturbations exciting the eigenvectors of the $K_{AB}$ subsystem neither oscillate nor grow, ensuring stability of these modes.

Similarly, the oscillation and stability conditions for the subsystem $A$, given in Equations \eqref{equ:AR_osc} and \eqref{equ:AR_stab}, depend on the eigenvalues of the matrices $R$ and $A^{-1}$.
If the matrix $A$ is SPD, these conditions indicate that the subsystem is overdamped and converges to the expected solution of the linear system of equations without oscillations. 
The convergence rate shown in Equation \eqref{equ:AR_conv} depends on the largest eigenvalue of the matrix $A^{-1}R$. 
This eigenvalue can be bounded as follows:
\begin{equation}
\lambda(A^{-1}R) \leq \frac{R_{max}}{\lambda_{min}(A)}
\end{equation}
where $R_{max}$ denotes the maximum value of the diagonal matrix $R$ and $\lambda_{min}(A)$ is the smallest eigenvalue of the matrix $A$. 
$R_{max}$ represents the maximum radius of the Gershgorin circles lying in the negative domain and may be considered as a measure of the maximum deviation from diagonal dominance.

\subsubsection{Domain of Positive Resistors}
\hspace*{\fill} \par
The proposed design ensures positive resistance in all elements, except those represented by the diagonal elements of the submatrix $K_B$. The domain of matrices that maintains strictly positive resistance can be identified as the domain of matrices that have nonpositive diagonals of $K_B$ after transformation. This condition is expressed as:
\begin{equation}
    K_{Bii} = D_{ii}-0.5(A_{ii}+|A_{ii}|)\leq 0
\label{equ:KBii}
\end{equation}
where substituting $D_{ii}$ from Equation \eqref{equ:Dii} yields:
\begin{equation}
    A_{ii} \geq (K_{s})_{ii} +\sum_{j=1,j\neq i}^{n} |A_{ji}|
\end{equation}
This condition indicates that the system $(A-K_s)$ must be diagonally dominant. 
The settling time of resistive networks is governed by the RC constants arising from parasitic capacitance in the network. 
However, the RC delay in resistive memristor arrays is in the order of tens of picoseconds \cite{7058989} and therefore can be neglected.
Consequently, solving a symmetric diagonally dominant (SDD) matrix with this design achieves a complexity $O(1)$, which is the theoretical maximum speed attainable for the solution process.

\begin{figure}[!t]  \centering
    \includegraphics[page=13,width=0.5\textwidth]{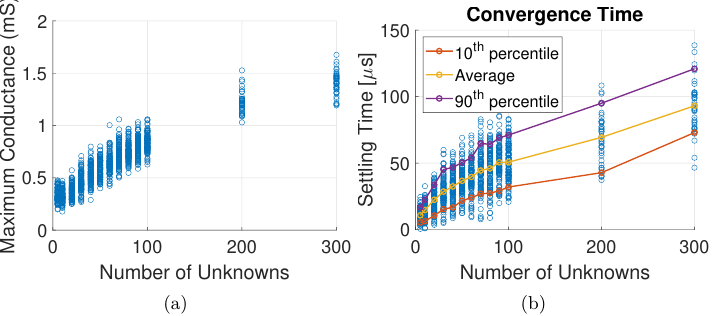}  
    \caption{Analysis of 1200 random dense systems (eigenvalues: [10 1000] $\mu S$) using the proposed design. Plot (a) shows the maximum conductance in the transformed system of each system. Plot (b) shows the maximum convergence time for each system }
    \label{fig:Complex2nd100} 
\end{figure}

\subsubsection{Bound Analysis}
\hspace*{\fill} \par
To gain deeper insight into the dynamics of the proposed circuit, upper and lower bounds on its settling time are derived. 
Since the governing differential equation is an overdamped second-order equation, adding a diagonal damping term increases the settling time and thus provides a valid upper bound. 
By adding the term $\left(2\tau K_N \dfrac{du}{dt}\right)$ to the governing Equation \eqref{equ:DE_2nd_order}, the differential equation reduces to the following
\begin{equation}    
2\tau^2\frac{d^2u}{dt^2}  +3\tau \frac{du}{dt}  = -M^{-1}(Ku-f)
\label{equ:DE_2nd_order_upper_bound}
\end{equation}
This upper-bound system remains a second-order differential equation that solves the preconditioned linear system $M^{-1}Ku=M^{-1}f$.

A lower bound on the settling time can be obtained by neglecting the delay introduced by the buffer OAs inside the negative resistance circuit. 
Applying Equation \eqref{equ:R_neg_DE_bound} for the transformed system and using the voltage relationship $u_j=-u_{n+j}$, the differential equation in vector form can be written as
\begin{equation}
2\tau \frac{du'}{dt}  = 3u - u'  \label{equ:DE_of_u'_bound}
\end{equation}
Substituting Equations \eqref{equ:du'dt} and \eqref{equ:3u-u'} in Equation \eqref{equ:DE_of_u'_bound} results in the following lower bound:
\begin{equation}
2\tau \frac{du}{dt}  = -M^{-1}(Ku-f)
\end{equation}
This lower bound is a first-order differential equation that also solves the same preconditioned system of equations $M^{-1}Ku=M^{-1}f$.

Both bounds demonstrate that the circuit dynamics are governed not by the original system $Ku=f$, but rather by its preconditioned form with $M^{-1}$ as the preconditioner.
For matrices close to diagonal dominance, $M^{-1}K$ is nearly the identity matrix, resulting in a more well-conditioned system that reduces error amplification.

\subsection{Complexity Studies for SPD Systems}\label{subsec:Des2Complex}

To evaluate the performance of the proposed design, complexity studies are performed for different classes of systems and compared to the feedback circuit in \cite{sun2019solving}.

\subsubsection{Spectrum-Constrained Random Matrices}\label{subsubsec:Des2RandMat}
\hspace*{\fill} \par

\begin{figure}[!t]  \centering
    \includegraphics[page=14,width=0.5\textwidth]{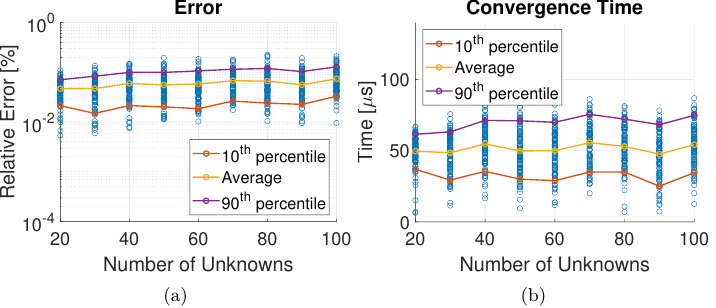}  
    \caption{Analysis of 900 systems (eigenvalues: [10 1000] $\mu S$) with maximum DD-deviation $R_{max}$ within 10\% of 800 $\mu S$ (density=1) using the proposed design. Plot (a) shows the maximum error between the converged solution and the theoretical values of each system. Plot (b) shows the maximum convergence time for each system }
    \label{fig:Complex2nd100K800} 
\end{figure}

The complexity studies in this section follow the same methodology used for the preliminary design studies. 
The objective is to validate that the computational complexity of the proposed design does not depend on the size of the system, but on matrix properties,specifically, the smallest eigenvalue and DD deviation. 

Random SPD but non-SDD matrices of various sizes were generated using MATLAB and then simulated in LTspice v.17.1.8 for both operating point and transient analyses. 
The OA model used is the manufacturer-supplied SPICE model for the AD712 which incorporates several nonideal behaviors, including finite bandwidth, finite slew rate, input offset voltage, bias currents, nonlinear output swing limits, and internal resistances. 
This more realistic OA model allows the study to capture circuit behavior beyond the simplified single-pole approximation. 

The random SPD systems generated in these studies eigenvalues ranges between and includes 10 $\mu S$ and 1000 $\mu S$.
For each randomly generated conductance matrix $A$, unknowns $x_i$ were randomly assigned values within the range [-0.5V,0.5V], and the corresponding vector $b$ was calculated by $b=Ax$.  
In the transient simulations, the supply voltage was modeled as a step function that changes from zero to $x_s^+$ or $x_s^-$ at 1 $\mu s$. 
This setup was used to analyze the settling time of the system and the accuracy of the solution for the unknown vector $x$.

\begin{figure}[!t]  \centering
    \includegraphics[page=15,width=0.5\textwidth]{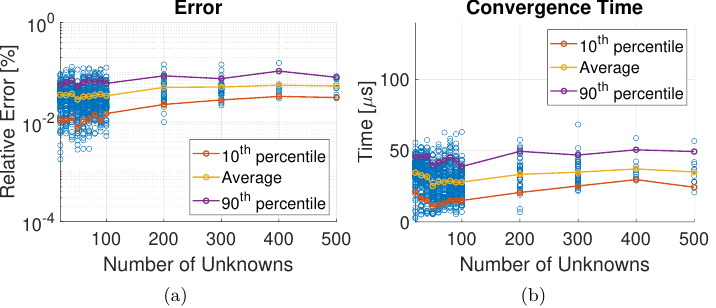}  
    \caption{Analysis of 1200 systems (eigenvalues: [10 1000] $\mu S$) with maximum DD-deviation within 10\% of 550 $\mu S$ (density=0.5) using the proposed design. Plot (a) shows the maximum error between the converged solution and the theoretical values of each system. Plot (b) shows the maximum convergence time for each system }
    \label{fig:Complex2nd50K550} 
\end{figure}

The first analysis focused on a set of full matrices that includes the 400 matrices previously analyzed in Section \ref{subsecDes1Tests} to compare the performance of the proposed design relative to that of the preliminary design. 
All simulated systems exhibited errors in the converged solution of less than 0.3$\%$. 
The results of the convergence time analysis are presented in Figure \ref{fig:Complex2nd100}.  
The proposed design demonstrated a remarkable improvement in convergence time, achieving at least a 100x speedup compared to the preliminary design. 
Systems that required milliseconds to solve using the preliminary design were solved in only tens of microseconds with the proposed design. 
Furthermore, the variation in convergence time across simulations was significantly reduced, highlighting the efficiency and consistency of the proposed design. 
The results in Figure \ref{fig:Complex2nd100} show a linear dependency between the settling/convergence time of the systems and the number of unknowns which can be attributed to the linear dependency between the number of unknowns and the DD-deviation $R_{max}$ represented by the maximum conductance in the transformed system $K$.

To further prove the independency of the complexity from the size of the system, a second test was conducted. 
In this test, the same procedure for generating random dense systems was applied, but only systems with a maximum DD-deviation $R_{max}$ or maximum conductance in the transformed system $K$ within 10\% of 800 $\mu S$ are selected. 
This procedure did not yield systems meeting this criterion for fewer than 20 or more than 100 unknowns. 
The results, presented in Figure \ref{fig:Complex2nd100K800}, show a constant trend in both error and settling time for solving these linear systems using the proposed design. 
On average, the error in the converged solution for 20x20 systems is 0.05\%, comparable to 0.07\% for 100x100 systems. 
Similarly, the average settling time remains nearly constant, ranging from 50$\mu s$ for 20x20 systems to 55 $\mu s$ for 100x100 systems.  

\begin{figure}[!t]  \centering
    \includegraphics[page=16,width=0.5\textwidth]{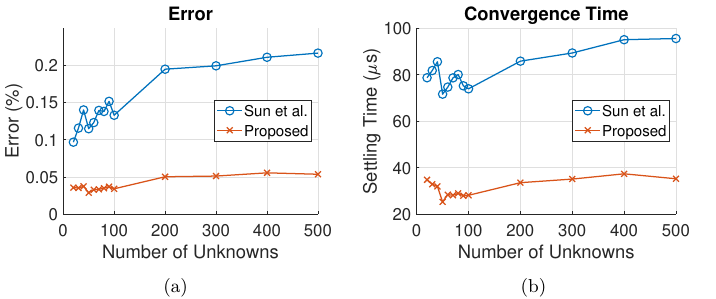}  
    \caption{Analysis of 1200 systems (eigenvalues: [10 1000] $\mu S$) with maximum DD-deviation within 10\% of 550 $\mu S$ (density=0.5) using the proposed design and the feedback design in Sun et al.  \cite{sun2019solving}. Plot (a) shows the average of the maximum error between the converged solution and the theoretical values of each system for both designs. Plot (b) shows the average of the maximum convergence time for each system for both designs.}
    \label{fig:Dir_vs_Eq_test5} 
\end{figure}

To verify this complexity observation in a wider range of unknowns, a third test was conducted. 
This test used a similar procedure to generate random systems with a density of 0.5 and a DD-deviation $R_{max}$ within 10\% of 550 $\mu S$. 
The results, shown in Figure \ref{fig:Complex2nd50K550}, again indicate a near-constant trend in both error and settling time. 
On average, the error in the converged solution for the 20x20 systems is 0.035\%, similar to 0.054\% for the 500x500 systems. 
Likewise, the average settling time is approximately 35$\mu s$ for both the 20x20 and the 500x500 systems, confirming that the settling time is independent of the number of unknowns.

The third test was conducted using the feedback circuit proposed by \cite{sun2019solving} to benchmark the performance of the proposed design. 
As shown in Figure \ref{fig:Dir_vs_Eq_test5}, the proposed equivalent-resistance design is on average twice as fast as the feedback design.
In addition, the proposed design reduces the error by a factor of two to four across the systems tested.

\subsubsection{Toeplitz Matrix}
\hspace*{\fill} \par
To benchmark the performance of the proposed design on structured matrices, the same Toeplitz matrix analyzed for the feedback architecture in \cite{sun2019fast} is reevaluated using the proposed equivalent design.
The Toeplitz matrix $T$ for different matrix size $n$ is defined as follows \cite{press2007numerical}:
\begin{equation}
T_{ij}=\frac{1}{\lvert i-j\rvert+1}
\end{equation}
The time complexity of solving the system $Tx=b$ is governed by the smallest eigenvalue of the matrix $T^{-1}R$, which is upper-bounded by $R_{max}/\lambda_{min}(T)$.
For this Toeplitz matrix, $\lambda_{min}(T)$ remains asymptotically constant as $n$ increases, whereas the maximum DD-deviation $R_{max}$ increases linearly with $\log(n)$. 
Consequently, the expected settling time scales as $O(\log(n))$, matching the complexity reported for the feedback design in \cite{sun2019fast}. 

A key distinction from the feedback circuit is that the proposed equivalent design is not dependent on uniform matrix scaling.
scaling the matrix by any constant multiplies both $R_{max}$ and $\lambda_{min}(T)$ by the same factor, leaving the ratio, and thus the settling time, unchanged. 
This property does not hold for feedback architectures.

\begin{figure}[!t]  \centering
    \includegraphics[page=17,width=0.5\textwidth]{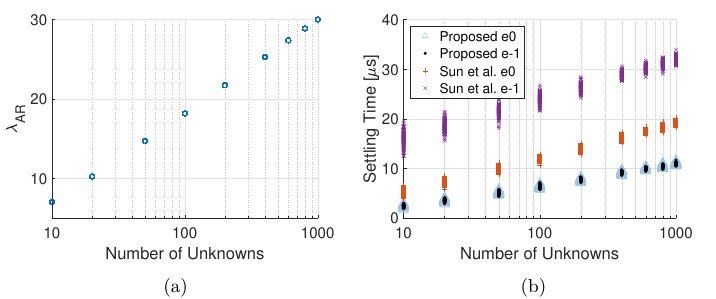}  
    \caption{(a) The minimum eigenvalue of the matrix $T^{-1}R$ for different sizes of the Toeplitz matrix $T$. (b) Settling time for solving the Toeplitz matrices scaled by 1 and 0.1 using the proposed Equivalent design and the Feedback design in \cite{sun2019fast}. The OAs used for both circuits have 10 MHz GBW. }
    \label{fig:Toeplitz_Test} 
\end{figure}

In practical implementations on memristor-arrays, this Toeplitz matrix require down-scaling because the matrix resistance values fall outside the feasible device range.
To evaluate this effect, both $T$ and its scaled version $0.1T$ were solved using the feedback design and the proposed equivalent design for multiple matrix sizes and multiple right-hand-side vectors $b$.
The results, summarized in Figure \ref{fig:Toeplitz_Test}, show that the proposed design achieves 2x speedup for the unscaled matrix $T$ and 3x for the scaled matrix $0.1T$ compared to the feedback architecture.
These results highlight the scaling resilience of the proposed architecture. 
This resilience provides more flexibility in the choice of the scaling factor to balance accuracy, power consumption, and hardware constraints without compromising the convergence speed


\subsubsection{Covariance Matrix}
\hspace*{\fill} \par

\begin{figure}[!t]  \centering
    \includegraphics[page=18,width=0.5\textwidth]{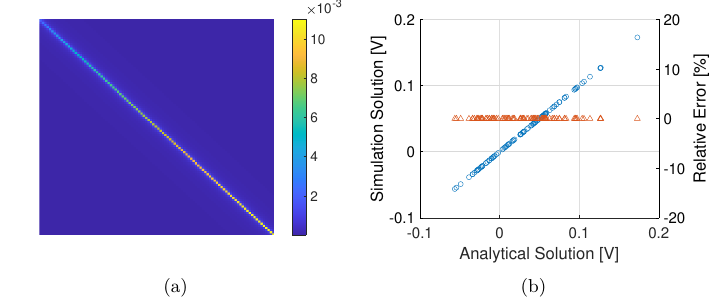}  
    \caption{(a) The 100x100 covariance matrix. (b) SPICE results for the 100 unknowns of solving the covariance matrix. Blue circles are associated with the left axis. The orange triangles are associated with the right axis, showing the relative error for each of the 100 variables.} 
    \label{fig:Covariance_Mat} 
\end{figure}

To evaluate the robustness of the proposed design to circuit nonidealities, specifically device variation and parasitic resistance, the same covariance matrix analyzed for the feedback architecture in \cite{sun2019solving} is scaled down by a factor of 1000 and reevaluated using the proposed equivalent design.

The unscaled 100x100 covariance matrix $C$ is defined as follows:
\begin{equation}
C_{ij} =
\begin{cases}
1+\sqrt{i} &  \text{if} \>\> i=j \\
\dfrac{1}{\lvert i-j \rvert} & \text{otherwise}  \\
\end{cases}
\end{equation}
The right-hand-side vector $b$ was randomly generated with values in the range $[0,1]$.
All simulations were performed on LTspice v.17.1.8. 
Using the exact resistance values, the proposed design accurately solved the system with errors below 0.001\% compared to the analytical solution as shown in Figure \ref{fig:Covariance_Mat}.

\begin{figure}[!t]  \centering
    \includegraphics[page=19,width=0.5\textwidth]{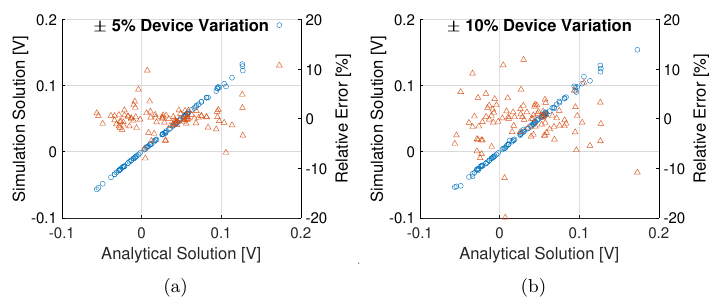}  
    \caption{SPICE simulation results of solving the 100x100 covariance matrix with random resistor device variations within (a) 5\% and (b) 10 \%. }
    \label{fig:Covariance_Device_Var} 
\end{figure}

The first nonideality examined is the device mismatch. 
In this test, each resistance value of the covariance matrix was randomly varied by $\pm5\%$ and $\pm10\%$. 
The results, shown in Figure \ref{fig:Covariance_Device_Var}, indicate that the solution error remained below 12\% in both cases except for two points whose analytical values are close to zero.  
Although the original covariance matrix has a condition number of 15.5, the error does not amplify significantly with device variation. 
This robustness of the design can be attributed to the fact that the circuit can be described as solving the preconditioned matrix $M^{-1}K$, which is close to the identity matrix for matrices that are nearly diagonal dominant. 
This lower condition number reduces the error amplification from device variation. 

\begin{figure}[!t]  \centering
    \includegraphics[page=20,width=0.5\textwidth]{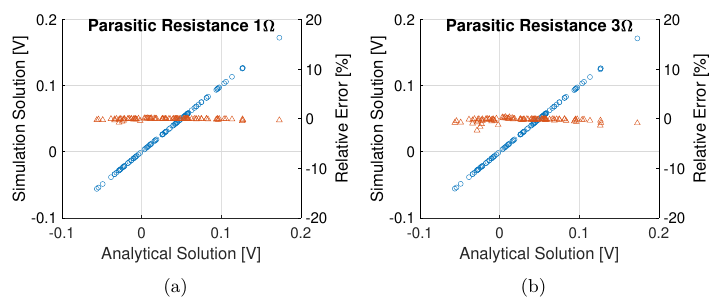}  
    \caption{SPICE simulation results of solving the 100x100 covariance matrix with parasitic interconnect resistance values of (a) 1$\Omega$ corresponding to 22 nm technology and (b) 3$\Omega$ corresponding to 18 nm technology. }
    \label{fig:Covariance_R_paras} 
\end{figure}

The second nonideality studied is parasitic wire resistance. 
In this study, the parasitic wire resistances between adjacent variable resistors are assumed to be 1 $\Omega$ corresponding to the 65 nm technology node and 3 $\Omega$ corresponding to the 22 nm technology node. 
Given that practical memristor arrays typically operate over a resistance range from hundreds of ohms to tens of kilo-ohms \cite{mem11bits}, the expected errors caused by parasitic resistance is minimal.
The results, shown in Figure \ref{fig:Covariance_R_paras}, confirm this expectation. 
Simulating the circuit including the parasitic resistance for the covariance matrix $C$ resulted in a max error of 0.8\% for the 1$\Omega$ case and 2.4\% for the 3$\Omega$ case.

Given that practical memristor arrays typically operate over a resistance range from hundreds of ohms to tens of kilo-ohms \cite{mem11bits}, the expected errors caused by parasitic resistance is minimal.
The results, shown in Figure \ref{fig:Covariance_R_paras}, confirm this expectation. 
Simulating the circuit including the parasitic resistance for the covariance matrix $C$ resulted in a max error of 0.8\% for the 1$\Omega$ case and 2.4\% for the 3$\Omega$ case.

\begin{figure}[!t]  \centering
    \includegraphics[page=21,width=0.5\textwidth]{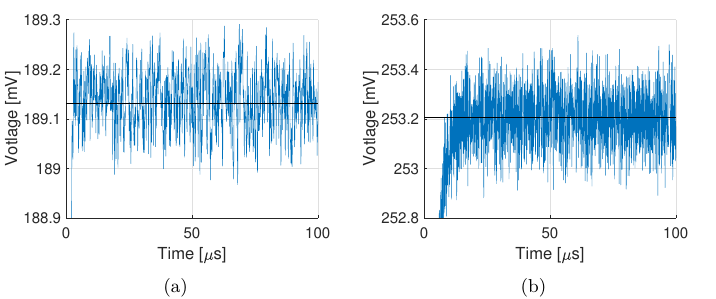}  
    \caption{SPICE simulation results for the transient analysis of thermal noise (bandwidth 1 GHz) of resistors when solving (a) the 100x100 covariance matrix (condition number: 15.6) and (b) a random 100x100 matrix with condition number of 100. }
    \label{fig:Covariance_noise} 
\end{figure}

The third nonideality investigated is device noise. 
In this study, the thermal noise of the variable resistors is modeled as white noise with a bandwidth of 1 GHz. 
The noise is modelled as a current source parallel to each resistor, using the standard thermal noise model \cite{OA_book}:
\begin{equation}
    \bar{I_n}^2=\frac{4kT\Delta f}{R}
\end{equation}
where $k$ is the Boltzmann constant, $T$ is the absolute temperature (set to 300 K), $R$ is the resistance value and $\Delta f$ is the noise bandwidth. 
The same transient analysis setup used in the previous studies is employed here, with the supply voltage applied as a step input at 1 $\mu s$.
In addition to the covariance matrix, a second 100x100 random SPD system with a condition number of 100 is also analyzed to assess noise sensitivity to condition number. 
The results, shown in Figure \ref{fig:Covariance_noise}, indicate that both systems converge reliably to their respective steady-state solutions despite the presence of thermal noise. 
In both cases, the resulting fluctuations in the output voltages remain bounded and do not exceed 0.13\% of the steady-state values, demonstrating the robustness of the proposed circuit to noise.

The observed robustness to thermal noise can be attributed to the combined effects of damping and implicit preconditioning in the circuit. 
The implicit preconditioning limits noise amplification, particularly for systems with higher condition number.
Meanwhile, the damping ensures that injected white noise appears only as small bounded fluctuations around the steady state solution rather than sustained deviations, guaranteeing stable convergence even under relatively large noise bandwidths.

\subsubsection{Operational Amplifier Specifications}

The accuracy and speed of the proposed design are influenced by the specifications of the operational amplifiers used. 
To examine the impact of amplifier choice, the same set of random matrices used in the preliminary design and test 1 of the complexity studies of the spectrum-constrained random systems were simulated using three different amplifiers. 
The first amplifier, AD712, was used in all previous nonideal simulations. 
The second amplifier, LTC2050, features a low input offset voltage to minimize errors in the converged solutions. 
The third amplifier, LTC6268, has relatively high slew rate and gain bandwidth product to reduce settling time. The main specifications of these amplifiers are summarized in Table \ref{tab:opamp_specs}.

\begin{table}[h]
\begin{center}
\caption{Operational Amplifiers Specifications}\label{tab:opamp_specs}%
\begin{tabular}{@{}llll@{}}
\hline
Specification                  & AD712             & LTC2050      &  LTC6268\\
\hline
Input Offset Voltage   & 1 $mV$            & 3 $\mu V$    &  2.5 $mV$   \\
Gain Bandwidth Product         & 4 MHz             & 3 MHz        & 500 MHz \\
Slew Rate                      & 20 $V/\mu s$      & 2 $V/\mu s$  & 400 $V/\mu s$ \\
\hline
\end{tabular}
\end{center}
\end{table}

\begin{figure}[!t]  \centering
    \includegraphics[page=22,width=0.5\textwidth]{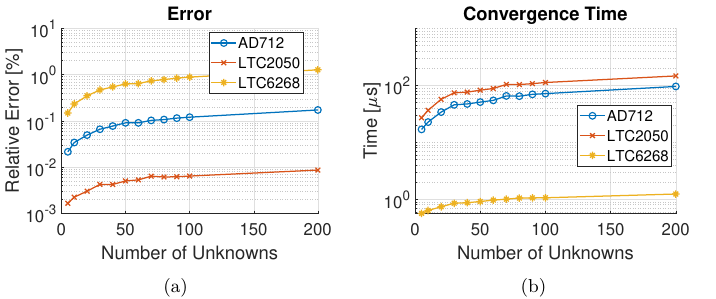}  
    \caption{Analysis of 1150 random systems (density=1) using the proposed design for different amplifiers. Plot (a) shows the 90\textsuperscript{th} percentile of the maximum error between the converged solution and the theoretical values of each system. Plot (b) shows the 90\textsuperscript{th} percentile of the maximum convergence time for each system }
    \label{fig:opamp_spec_results} 
\end{figure}

The simulation results, presented in Figure \ref{fig:opamp_spec_results}, reveal a tradeoff between speed and accuracy. 
Using the LTC2050 amplifier reduces the error in the converged solution by an order of magnitude compared to the AD712 but increases the settling time by approximately 1.6 times. 
Conversely, the LTC6268 amplifier significantly speeds up the system, achieving a two-orders-of-magnitude reduction in settling time, but with an increase in error by an order of magnitude. 
These highlight that a lower input offset voltage results in higher accuracy in the converged solution, while a higher gain bandwidth product leads to faster system speed. 

The selection of the OAs depend on the performance requirements of the target application.
For applications such as high-fidelity simulations, where numerical accuracy is prioritized over computational speed, OAs with very low input offset voltage are preferred.
On the other hand, for real-time simulation applications, where higher error tolerance is acceptable in exchange for faster convergence, OAs with high gain–bandwidth product are more suitable.

\section{Discussion} \label{sec:disc}

This article presents two analog circuit designs, preliminary and proposed, aimed at accelerating the solution of symmetric linear systems of equations, an essential computational task in various applications. Although both designs leverage the analogy between 1D spring systems and resistive networks, the proposed design improves the process by reducing the number of circuit components by approximately 70\% compared to the preliminary design, despite adding $n$ more unknowns to the system. This reduction in components is detailed in Table \ref{tab:n_comp}.

\begin{table}[h]
\begin{center}
\caption{Number of Circuit Components}\label{tab:n_comp}%
\begin{tabular}{@{}llll@{}}
\hline
Number of                  & Preliminary Design       & Proposed Design   \\ \hline
Unknowns                   & $n$            & $2n$       \\
Variable Resistors         & $n^2 +2n$      & $2n^2 +1$  \\
Resistors ($10\ k\Omega$)  & $2(n^2 +n)$    & $4n$       \\
Analog Switches            & $1.5n^2 +2.5n$ & $3n$       \\
OpAmps                     & $2(n^2 +n)$    & $4n$       \\
\hline
\end{tabular}
\end{center}
\end{table}

The proposed design is tailored specifically for solving Symmetric Positive Definite (SPD) systems, which define its stable domain. Within this domain, the performance of the system varies based on the matrix properties:
\begin{itemize}
    \item \textbf{Diagonally Dominant SPD Systems:} For these systems, the solution is achieved with $O(1)$ complexity, irrespective of the matrix size or specific properties, since the circuit consists exclusively of resistors. 
    This enables the system to reach the theoretical maximum speed.
    \item \textbf{Non-Diagonally Dominant SPD Systems:} The accuracy and speed of the solution depend on the specifications of the OAs used and the system matrix properties, particularly, the smallest eigenvalue and the maximum deviation from diagonal dominance. More importantly, the solution speed remains independent of the matrix size.
\end{itemize}

The reliance on OAs is a widely recognized bottleneck in analog computing architectures, including feedback-based solvers reported in the literature. 
Power consumption and large area requirements fundamentally limit the feasible number of OAs in a single array, regardless of the specific circuit topology.

The proposed transformation represents a step toward mitigating this bottleneck by reducing the required number of active elements.
Instead of scaling with the number of unknowns $n$, the transformation enables the number of active elements to scale with the number of non-diagonally dominant columns $c$.
Consequently, an $n \times n$ problem can be solved using at most $4c$ OAs, where $c \ll n$ for matrices that are close to diagonal dominance, such as those arising in finite element solvers for dynamic solid mechanics simulations. 
The SPICE simulations for systems with a large number of unknowns presented in this work are therefore primarily intended to illustrate convergence and scaling behavior. 
More advanced matrix transformations that build on the framework introduced here offer a promising path toward further reducing the number of required active elements.

While the proposed design achieves size-independent convergence during the analog solution phase, it requires a preprocessing step to compute the sum of the absolute values of the columns, 
which has a worst-case digital complexity of $O(n^2)$. 
Several approaches can be used to perform this preprocessing, with the choice depending on whether physical area or computational speed is prioritized:
\begin{itemize}
    \item \textbf{System Assembly:} In applications such as the Finite Element Method for computational solid mechanics, the calculation of column sums can be naturally incorporated into the system assembly process. 
    During element-wise assembly, the required sums can be accumulated along with the matrix entries with negligible additional overhead. Thus, the preprocessing complexity is still governed by the system assembly complexity rather than the extra column sum operation. 
    \item \textbf{Parallel Resistors:} Diagonal entries of the submatrix $K_B$ can be implemented by representing each column using individual variable resistors connected in parallel, each programmed to the conductance of the corresponding matrix element. 
    This approach eliminates the need for an explicit computational step at the expense of an increase in physical area, which requires $n^2$ additional programmable resistive elements.
    \item \textbf{Matrix Vector Multiplication (MVM):} An additional analog MVM operation can be used to multiply the matrix by a vector of ones. 
    This operation can be executed using the same crossbar hardware and incurs a constant-time $O(1)$ analog cost.
\end{itemize}
These alternatives highlight a fundamental design trade-off between circuit area and computational speed, 
while preserving the size-independent convergence behavior of the proposed analog solver.


\section{Conclusion} \label{sec:conc}

The presented research work introduces a fundamentally new class of analog compute-in-memory (CIM) solvers, resistive network mapping, that directly maps the given SPD system of equations into a resistive network. 
This mapping is only realized after leveraging the noninverting OA configuration with some modifications to map negative conductance.
Another key aspect of the architecture is the proposed transformation of the original $n\times n$ system (up to $(n^2-n)/2$ negative conductance values) into an equivalent $2n\times 2n$ system (up to $n$ negative conductance values). 
Mapping and solving this transformed system reduces the needed component count by more than 70\%, consumes less power, and speeds up the computation by orders of magnitude without additional computational overhead.

For the subclass of symmetric diagonally dominant (SDD) matrices, mapping the transformed system results in a purely passive resistive network whose nodal voltages settle in effectively zero time. 
This yields true $O(1)$ complexity where the solution speed reaches the theoretical speed limit for analog-CIM architectures and outperforms existing analog solvers, whose settling times are limited by RC constants and feedback loop dynamics, by orders of magnitude. 
Compared to conventional digital algorithms (even highly optimized sparse solvers), the passive resistive network achieves multiple orders of magnitude speedups for large SDD systems, while consuming a fraction of the energy.

More generally, for any SPD system, the solution speed remains independent of the number of unknowns and depends only on other matrix properties, particularly, the smallest eigenvalue and the maximum deviation from diagonal dominance.
The resulting circuit dynamics is shown to correspond to an overdamped second-order differential equation that implicitly solves a preconditioned system, significantly improving numerical conditioning relative to the original problem. 
Numerical studies on random spectrum-constrained systems confirm the derived complexity bounds, demonstrating that convergence depends on matrix properties rather than the number of unknowns. 
In addition, the number of active components scales with the number of non–diagonally dominant columns and is therefore always bounded by $n$, further decoupling hardware complexity from problem size.

Benchmarking against the feedback architecture in \cite{sun2019solving} highlights the advantages of the proposed equivalent design.
For both random SPD systems with constrained spectra and structured Toeplitz matrices, the proposed solver consistently achieves faster convergence. 
Unlike feedback-based designs, the settling time of the proposed circuit is invariant to the uniform scaling of the matrix, providing greater flexibility in selecting resistance ranges to balance accuracy, dynamic range, power consumption, and hardware constraints. 

The robustness of the architecture is validated through nonideality analyses. 
SPICE simulations incorporating realistic OA models, device variations, parasitic resistance, and thermal noise demonstrate stable convergence. The implicit preconditioning inherent to the architecture reduces the sensitivity to component tolerances for matrices with high condition numbers, while the overdamped dynamics suppress noise-induced deviations, ensuring reliable operation under nonideal conditions.

The accuracy of the proposed design is limited by the specifications of the variable resistors and OAs used, which will keep improving with the rapid improvements in device technology. 
A notable advantage of the proposed design is its compatibility with the cross-point architecture, commonly used in resistive memory arrays. 
This compatibility facilitates in-memory computations, enabling multiple operations such as MVM within the same framework. 

In summary, we show that resistive network mapping has high potential in computational linear algebra by eliminating the mathematical process of solving linear systems of equations. 
The proposed design offers substantial improvements in speed and efficiency, particularly for SPD systems, such as those arising from Finite Element Analysis in Computational Solid Mechanics. 
Future work could explore extending this design to a broader range of linear systems and developing more robust solutions for handling non-positive definite matrices efficiently.

\section*{Acknowledgments}

Aspects of the technology described in this paper are the subject of U.S. Provisional Patent Application No. 63/690,322, filed on September 4, 2024, and U.S. Provisional Patent Application No. 63/839,979, filed on July 8, 2025.


{\appendices

\section{Poles Derivation}

Definitions
\begin{equation}
K =\begin{bmatrix}K_A & K_B \\ K_B & K_A \end{bmatrix}
\end{equation}
\begin{equation}
K_{AB} = K_A+K_B
\end{equation}
\begin{equation}
R = \text{diag}_+(K_B)
\end{equation}
\begin{equation}
K_A -K_B = A
\end{equation}
\begin{equation}
K_N = \begin{bmatrix}R & 0 \\ 0 & R \end{bmatrix}
\end{equation}
\begin{equation}
N = \begin{bmatrix}R & -R \\ -R & R \end{bmatrix}
\end{equation}
\begin{equation}
Q = \frac{1}{\sqrt{2}}\begin{bmatrix}I & I \\ -I & I \end{bmatrix}
\end{equation}
\begin{equation}
P = K+N
\end{equation}
\begin{equation}
M = P+K_N
\end{equation}

Calculations from these Definitions:
\begin{align}
QKQ^T &= \begin{bmatrix}K_{AB} & 0 \\ 0 & A \end{bmatrix} \\
QNQ^T &= \begin{bmatrix}0 & 0 \\ 0 & 2R \end{bmatrix} \\
QK_NQ^T &= \begin{bmatrix}R & 0 \\ 0 & R \end{bmatrix} \\
QMQ^T &= Q(K+N+K_N)Q^T \\
QMQ^T &= \begin{bmatrix}K_{AB}+R & 0 \\ 0 & A+3R \end{bmatrix} \\
Q(3M-2K_N)Q^T &= \begin{bmatrix}3K_{AB}+R & 0 \\ 0 & 3A+7R \end{bmatrix}
\end{align}
Governing DE
\begin{equation}    
2\tau^2 M \frac{d^2u}{dt^2}  +3\tau M\frac{du}{dt} -2\tau K_N\frac{du}{dt}  = -(Ku-f)
\end{equation}
To define the system poles:
\begin{equation}
\det\left(2\tau^2Ms^2+\tau(3M-2K_N) s+K \right) = 0
\end{equation}
Applying the orthogonal transformation:
\begin{equation}
\det\left(2\tau^2s^2QMQ^T
+\tau s Q(3M-2K_N)Q^T 
+QKQ^T \right) = 0
\end{equation}

\subsection{The first $n$ equations:}
\begin{equation}
\det\left(2\tau^2s^2(K_{AB}+R)+\tau s(3K_{AB}+R) +K_{AB} \right) = 0
\end{equation}
Multiply by $\det\left(K_{AB}^{-1}\right)$:
\begin{equation}
\det\left(2\tau^2s^2(I+K_{AB}^{-1}R)+\tau s(3I+K_{AB}^{-1}R) +I \right) = 0
\end{equation}
Left multiply by eigenvectors of $K_{AB}^{-1}R$ defined as $\Phi_{KR}$ and right multiply by its transpose $\Phi_{KR}^T$
\begin{equation}
\det\left(2\tau^2s^2(I+\Lambda_{KR})+\tau s(3I+\Lambda_{KR}) +I \right) = 0
\end{equation}
The equations are then decoupled for the individual eigenvalues of $K_{AB}^{-1}R$ defined as $\lambda_{KR}$ as follows:
\begin{equation}
2\tau^2s^2(\lambda_{KR}+1)+\tau s(\lambda_{KR}+3) +1 = 0
\end{equation}
Oscillation Check: $b^2-4ac$:
\begin{equation}
\tau^2(\lambda_{KR}+3)^2 - 4(2\tau^2)(\lambda_{KR}+1)
\end{equation}
This formula simplifies to the following expression:
\begin{equation}
\tau^2(\lambda_{KR}-1)^2 \label{equ:K_AB_osc}
\end{equation}
The resulting expression is always positive, causing no oscillations in the response of vectors that excite the eigenvectors from the system $K_{AB}$.\\
Convergence/Stability Condition:
\begin{equation}
\frac{-b\pm \sqrt{b^2-4ac}}{2a}
\end{equation}
\begin{equation}
-\frac{1}{4\tau}\frac{ (\lambda_{KR}+3)\pm(\lambda_{KR}-1)}{(\lambda_{KR}+1)}
\end{equation}
The stability/convergence of this subsystem is achieved in the domain where the above expression is negative. 
This condition simplifies to the following condition:
\begin{equation}
\lambda_{KR}>-1 \label{equ:K_AB_stab}
\end{equation}

\subsection{The second $n$ equations:}
\begin{equation}
\det\left(2\tau^2s^2(A+3R)+\tau s(3A+7R) +A \right) = 0
\end{equation}
Multiply by $\det\left(A^{-1}\right)$:
\begin{equation}
\det\left(2\tau^2s^2(I+3A^{-1}R)+\tau s(3I+7A^{-1}R) +I \right) = 0
\end{equation}
Left multiply by eigenvectors of $A^{-1}R$ defined as $\Phi_{AR}$ and right multiply by its transpose $\Phi_{AR}^T$
\begin{equation}
\det\left(2\tau^2s^2(I+3\Lambda_{AR})+\tau s(3I+7\Lambda_{AR}) +I \right) = 0
\end{equation}
The equations are then decoupled for the individual eigenvalues of $A^{-1}R$ defined as $\lambda_{AR}$ as follows:
\begin{equation}
2\tau^2s^2(3\lambda_{AR}+1)+\tau s(7\lambda_{AR}+3) +1 = 0
\end{equation}

Oscillation Condition: $b^2-4ac <0$:
\begin{equation}
\tau^2(7\lambda_{AR}+3)^2 - 4(2\tau^2)(3\lambda_{AR}+1)
\end{equation}
\begin{equation}
\tau^2(49\lambda_{AR}^2+18\lambda_{AR}+1)>0
\end{equation}
\begin{equation}
-0.3<\lambda_{AR}<-0.068 \label{equ:AR_osc}
\end{equation}

Convergence/Stability Condition:
\begin{equation}
\frac{-b\pm \sqrt{b^2-4ac}}{2a} <0
\end{equation}
\begin{equation}
\frac{-1}{\tau}\frac{ (7\lambda_{AR}+3)\pm\sqrt{49\lambda_{AR}^2+18\lambda_{AR}+1}}{4(3\lambda_{AR}+1)}
 < 0
\end{equation}
By plotting: 
\begin{equation}
\lambda_{AR}>-1/3 \label{equ:AR_stab}
\end{equation}

Convergence Rate:
\begin{equation}
\max\left(\frac{-b\pm \sqrt{b^2-4ac}}{2a} \right) 
\end{equation}
Approximately:
\begin{equation}
b^2-4ac \approx \tau^2(7\lambda_{AR}+1)^2
\end{equation}
\begin{equation}
\frac{-\tau(7\lambda_{AR}+3) + \tau(7\lambda_{AR}+1)}{4\tau^2(3\lambda_{AR}+1)}  
\end{equation}
\begin{equation}
\frac{-1}{\tau(6\lambda_{AR}+2)}  \label{equ:AR_conv}
\end{equation}

}

%

\bibliographystyle{IEEEtran}
\bibliography{myreferences}

\vfill

\end{document}